\documentclass{article}
\usepackage{arxiv}
\usepackage[utf8]{inputenc} 
\usepackage[T1]{fontenc}    
\usepackage{hyperref}       
\usepackage{url}            
\usepackage{booktabs}       
\usepackage{amsfonts}       
\usepackage{nicefrac}       
\usepackage{microtype}      
\usepackage{lipsum}         
\usepackage{graphicx}
\usepackage{natbib}
\usepackage{doi}

\usepackage{mathtools}
\usepackage{amsmath}
\usepackage{bm}
\usepackage{multirow}
\usepackage{subcaption}
\usepackage{esint}
\usepackage{cleveref}       
\title{On nonlinear transitions, minimal seeds and exact solutions for the geodynamo}

\newcommand{\vect}[1]{\boldsymbol{#1}}
\newcommand{\matr}[1]{\mathbf{#1}}

\usepackage{authblk}

\author[1]{%
	\href{https://orcid.org/0000-0003-0994-2013}{Calum S. Skene}\thanks{Corresponding author: \texttt{c.s.skene@leeds.ac.uk}}%
}
\author[2]{%
	\href{https://orcid.org/0000-0003-4739-7466}{Florence Marcotte}%
}
\author[1]{%
	\href{https://orcid.org/0000-0003-0205-7716}{Steven M. Tobias}%
}
\affil[1]{Department of Applied Mathematics, University of Leeds, Leeds LS2 9JT, UK}
\affil[2]{Universit\'e C\^ote d'Azur, Inria, CNRS, LJAD, France}

\renewcommand{\shorttitle}{On nonlinear transitions, minimal seeds and exact solutions for the geodynamo}

\hypersetup{
pdftitle={On nonlinear transitions, minimal seeds and exact solutions for the geodynamo},
pdfsubject={},
pdfauthor={Calum S. Skene, Florence Marcotte, Steven M. Tobias},
pdfkeywords={},
}

\begin{document}
\maketitle

\begin{abstract}
Nearly fifty years ago, \cite{roberts1978magneto} postulated that Earth's magnetic field, which is generated by turbulent motions of liquid metal in its outer core, likely results from a subcritical (finite-amplitude) dynamo instability characterised by a dominant balance between Coriolis, pressure and Lorentz forces. Here we numerically explore subcritical convective dynamo action in a spherical shell, using techniques from optimal control and dynamical systems theory to uncover the nonlinear dynamics of magnetic field generation. Through nonlinear optimisation, via direct-adjoint looping, we identify the minimal seed --- the smallest magnetic field that attracts to a nonlinear dynamo solution. Additionally, using the Newton-hookstep algorithm, we converge stable and unstable travelling wave solutions to the governing equations. By combining these two techniques, complex nonlinear pathways between attracting states are revealed, providing insight into a potential subcritical origin of the geodynamo. This paper showcases these methods on the widely studied benchmark of \citet{Christensen2001}, laying the foundations for future studies in more extreme and realistic parameter regimes. We show that the minimal seed reaches a nonlinear dynamo solution by first approaching an unstable travelling wave solution, which acts as an edge state separating a hydrodynamic solution from a magnetohydrodynamic one. Furthermore, by carefully examining the choice of cost functional, we establish a robust optimisation procedure that can systematically locate dynamo solutions on short time horizons with no prior knowledge of its structure.
\end{abstract}

\keywords{geodynamo \and magnetoconvection \and  subcritical transition \and adjoint-based optimisation}

\section{Introduction}

\subsection{Balances in the geodynamo: weak and strong field branches}

The problem of generation of Earth's magnetic field is an important and challenging one for geophysics; the magnetic field plays a key role in deflecting the solar wind and therefore is important in helping to protect us from the harmful effects of the charged particles and the depletion of ozone leading to increased exposure to harmful ultraviolet radiation \citep[see e.g.][]{arsenovicetal2024}. The geomagnetic field is predominantly dipolar, and so large scale, with a dipole axis that is offset from the axis of rotation. It has variability on a wide range of timescales from years to centuries and includes such dynamical features as excursions and reversals \citep[see e.g.][]{CONSTABLE2006, Roberts2008}.

Earth's magnetic field is believed to be generated by dynamo action in its fluid outer core where convective motions (either compositionally or thermally driven) in the electrically conducting medium can drive currents and hence magnetic fields via induction \citep[see e.g.][]{Moffatt_Dormy_2019}. The geodynamo problem involves the simultaneous solution of the partial differential equations governing the evolution of the fluid (the momentum equation), the magnetic field (the induction equation) and the evolution equation for the co-density. The theoretical challenge arises owing to the vast range of temporal scales (and to a lesser extent spatial scales) that need to be resolved in the momentum equation (with the induction and codensity - a variable which combines the buoyancy effects of temperature and chemical species concentration \citep{Braginsky_1995} - equations causing fewer problems); the large range of temporal scales leads to the requirement that the non-dimensional equations be solved at extreme parameter values and hence the impossibility of direct solution via computational methods.

Nonetheless progress has been made on a number of fronts. Heroic computations \citep[following on from the pioneering work of][]{Glatzmaier_1995} are starting to achieve small enough parameters (though still extremely far from their true values) that interesting dynamics are achieved \citep[see e.g.][]{agf2017,Schaeffer_2017,sga19}. This computational endeavour has been predicated on the understanding that it is important to approach the correct regime by keeping the balances in the momentum equation and codensity equations realistic as one increases the separation of timescales. This is achieved by considering a \textit{distinguished limit}, as suggested by \citet{dormy_2016}, so that all the non-dimensional parameters scale together to achieve the correct balance.

What is the correct balance for the geodynamo? It is widely thought that the magnetic field in the Earth's core is strong enough to play a significant role in the dynamics of the flow; at least at some lengthscale. For a detailed discussion of how, and at what scale, the magnetic Lorentz force may enter in the dynamics, see \citet{aurnou_king_2017} or \citet{tobias_2021}. 
Theoretical support for the importance of the magnetic field arises from theoretical and computational studies of magnetoconvection \citep[see e.g.][]{Proctor_1994,Fearn_1998}. \citet{roberts1978magneto} recognised that there may be two branches of the Earth's dynamo; a weak field branch and a strong field branch. The weak field branch arises in a traditional bifurcation scenario; as the thermal driving (as measured by the non-dimensional Rayleigh number) is increased hydrodynamic thermal convection sets in. Because the Earth is a rapid rotator, this convection sets in at very small length-scale and at very high Rayleigh number. Eventually the convection becomes strong enough so that the magnetic Reynolds number of the flow is large enough for dynamo action to onset. This leads to the formation of weak multipolar magnetic fields \citep{Petitdemange}; here these fields are weak in the sense that their magnetic energy is equivalent to, or less than, the kinetic energy of the flow. 

A more efficient situation occurs when the magnetic field is strong. The strong magnetic field breaks the restriction imposed on the flow by the fast rotation and leads to the onset of convection on larger lengthscales at more moderate values of the Rayleigh number measuring thermal driving. Thus one might expect to find a more efficient \textit{subcritical strong field branch} for values of the Rayleigh number well below onset of weak field dynamo action (and potentially even below the onset of hydrodynamic convection). On the strong field branch the magnetic energy should be much larger than the kinetic energy of the flow and the Lorentz force may even be strong enough to enter into the leading order balance, competing with the Coriolis force and the pressure gradient to give a magnetostrophic balance (at least at certain scales \citep{dormy_2016}). The magnetic field in this situation acts so as to make the dynamo behave in a more efficient manner --- an example of an essentially nonlinear dynamo \citep{tcb:2011}. There is therefore a strong theoretical drive to understand subcritical dynamo action in the context of rapidly rotating spherical shells.

\subsection{Computing essentially nonlinear dynamos}

It is, however, non-trivial to compute nonlinear dynamos in the subcritical regime. Although timestepping methods are able sometimes to  locate these, the results are often haphazard and sometimes owe their success to a slice of luck. A more systematic approach relies on the computation of fixed points. The latter builds on quasi-Newton iterative solvers to compute equilibrium dynamo solutions by continuation while progressing through the parameter space. Such solvers however require fairly good first guesses for the procedure to converge toward a solution. This usually implies finding a starting point in the linearly unstable regime (whenever possible) and then following the saturated solution down as the control parameter is gradually decreased toward the linearly stable regime. Alternatively, one may turn the subcritical problem into a supercritical one e.g. through the addition of a suitable physical forcing, which is then gradually removed \citep{Waleffe_2003} --- in a dynamo context, this method was successfully used by \citet{Rincon_2007} to identify subcritical equilibrium solutions in rotating, magnetohydrodynamic (MHD) plane Couette flow. Moreover, while this continuation approach has the considerable advantage of computing unstable as well as stable equilibria, it can only compute exact solutions that are travelling waves or at least recurring states, which limits its applicability. 

In the context of wall-bounded turbulence, the objective of such computations of equilibria is not to find the attracting, statistically steady turbulent state. This state is fairly easy to reproduce numerically (or indeed experimentally). The theoretical challenge rather lies with the identification of unstable coherent states and their dynamical role in the transition to turbulence \citep{toh2003,Schneider_2007,wang2007,khapko2016}, with the control of the transition as a long-term objective. In the context of subcritical dynamo problems, however, identifying the attracting dynamo state in itself can be very uncertain. It is imperative that the computed dynamo states be allowed to be highly fluctuating. Furthermore, it is desirable that as little knowledge of the dynamo mechanism as possible be required for the computation. To overcome these limitations, a robust way to numerically identify an elusive subcritical state without prior guesses of the equilibrium is through a combination of short-time optimal control and long-time direct numerical simulations, as demonstrated by \citet{Mannix2022a} with nonlinear dynamos. 

This alternative approach builds on the mathematical framework of transient growth and nonmodal stability analysis (see for example \citet{Schmid_2007,Kerswell_2018} for reviews of both linear and nonlinear nonmodal stability analysis, respectively). It has been widely used in the last decade to compute `minimal' (i.e. least-energy) perturbations that nonlinearly trigger transition to hydrodynamic turbulence in shear flows \citep{Pringle_2010,Duguet_2010, cherubini2010,Vavaliaris_2020}, which no alternative methods can currently compute. In particular, fixed point methods cannot identify minimal seeds since these are not steady or recurring solutions of the underlying dynamical system. The review of \citet{kerswell2014} illustrates this approach for identifying minimal seeds in dynamical systems in general, and identifies three key considerations. The first outlines the importance of the optimisation time horizon, which must be large enough to overcome initial transients but not so large as to render convergence difficult due to the increasing sensitivity of the optimum to the initial condition. Secondly, including nonlinearities is imperative since the most-amplified perturbations in the linear problem are (generally) not the minimal seeds for subcritical transition. Lastly, the functional to be optimised to compute least-energy perturbations need not be the energy. Further to these points, \citet{Mannix2022a} found that the spatial structure of minimal dynamo seeds bears no resemblance to that of the subcritical, nonlinear state -- a finding that is consistent with studies of minimally-triggered turbulence in shear flows -- and that the sequence of events they trigger is revealing of the mechanisms of subcritical dynamo action in the considered systems.

\subsection{Triggering nonlinear transition to subcritical, convective dynamos}

In this paper, we investigate the geodynamo benchmark of \citet{Christensen2001} as an ideal starting point for developing the methodology needed to identify minimal seeds that lead to subcritical dynamo action in a geodynamo setup. This benchmark study contains three cases, of which we consider the first two; a purely hydrodynamic benchmark and a subcritical dynamo benchmark. In the hydrodynamic benchmark, convective instability gives rise to Taylor columns with $m=4$ symmetry that ultimately persists as a travelling wave. Although at the benchmark parameters the conducting (steady base-state) is unstable to convection, the resulting convecting state is linearly stable to dynamo action. In this manner, the form of the magnetic field initial condition is crucial, and needs to have a large enough amplitude as well as the correct structure in order to reach a nonlinear dynamo.

Although originally intended as a benchmark for Boussinesq MHD spherical shell codes, the `modest' benchmark parameters provide rich dynamical behaviour that has subsequently been studied. The hydrodynamic bifurcation behaviour has been explored by \citet{Feudel2015}, which shows that there are multiple stable nonlinear hydrodynamic travelling waves present at the benchmark parameters. Subsequently, \citet{Feudel2017} studied the bifurcation nature of nonlinear dynamo solutions possible in the system, showing that at the benchmark parameters there are stable and unstable MHD states that arise via a saddle node bifurcation, and that the MHD solution branch is smoothly connected to a purely hydrodynamic branch where a secondary Hopf bifurcation occurs. In these studies, the nonlinear travelling waves were converged using a matrix-free Newton solver using implicit integration of the linear terms  as a preconditioner \citep{Mamun_1995}. More recently, \citet{skene2024} used a weakly nonlinear analysis to study the saturation of the convective instability in the purely hydrodynamic case near the onset of convection.

While these studies illuminate the dynamical states possible for the system, they do not provide information on the complex nonlinear pathways between these states or their basins of attraction. A more systematic programme for mapping out the landscape of subcritical dynamos is needed, with tools developed for uncovering dynamics in a spherical shell. It is the first steps of just such a programme that we shall describe in this paper, which combines nonlinear optimisation with the identification of the unstable and stable travelling waves present in the system. Optimal control, on the one hand, can identify the magnetic field with the smallest energy that attracts to a dynamo solution. Variational optimisation in dynamo theory dates back to \citet{backus1958}, who derived a bound on the magnetic Reynolds number required for dynamo action in a sphere. More recently, optimisation of steady velocity fields has been performed in various geometries to maximise linear dynamo growth rates \citep{willis2012,Chen2015,chen_herreman_li_livermore_luo_jackson_2018,herreman2018}. In the present paper we use optimal control on the full set of MHD equations to provide a limit on the basins of attraction for reaching a nonlinear dynamo. By considering the cost functional to extremise, we furnish a robust procedure that can identify dynamos using short time horizon optimisations, enabling this procedure to be scaled up in future studies. On the other hand, we also use a Newton-hookstep method \citep{Dennis_1996} to find stable and unstable travelling waves in the system, and examine them in light of the dynamical pathways between important states. We demonstrate that the Newton-hookstep procedure is able to converge unstable magnetic states without preconditioning, and therefore provide the foundation for future studies where these exact solutions will lie within a turbulent and chaotic flow field. The structure of the paper is as follows; the mathematical and numerical setup is outlined in \S \ref{sec:mathematics}, the results are detailed in \S \ref{sec:results}, and conclusions are offered in section \S \ref{sec:conclusions}.

\section{Mathematical and numerical setup} \label{sec:mathematics}
\subsection{Governing equations}

\begin{figure}
	\centering
	\includegraphics[scale=0.8]{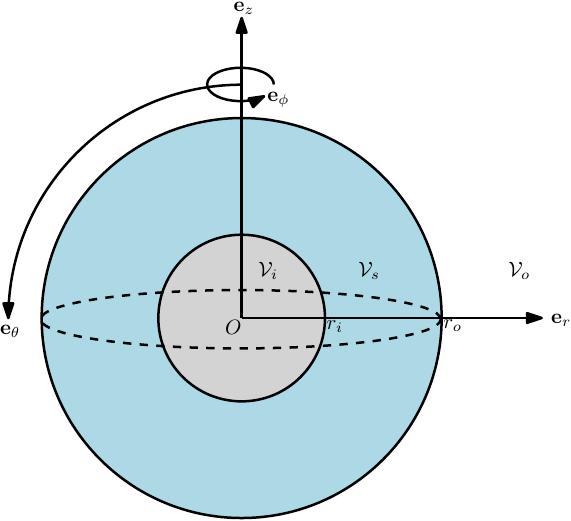}
	\caption{Sketch of the numerical domain. The fluid domain $\mathcal{V}_s$ (in blue) is surrounded by two insulating, solid domains: the inner core $\mathcal{V}_i$ (in grey) and an outer mantle $\mathcal{V}_o$.}
	\label{fig:domain}
\end{figure}

We model the geodynamo by considering the setup illustrated in figure \ref{fig:domain}. Spherical coordinates $(\theta,\phi,r)$ are used where $\theta$, $\phi$, and $r$ are the colatitudinal, azimuthal, and radial coordinates, respectively. Our domain $\mathcal{V}=\mathbb{R}^3$ is split into three parts, i.e. $\mathcal{V}=\mathcal{V}_i\cup\mathcal{V}_s\cup\mathcal{V}_o$. The spherical shell region $\mathcal{V}_s$, where $r_i<r<r_o$, is filled with a conducting fluid. The inner region $\mathcal{V}_i$ ($r<r_i$) and outer region $\mathcal{V}_o$ ($r>r_o$) do not consist of a fluid, but can support a magnetic field and therefore are needed to complete our geodynamo description: the inner region $\mathcal{V}_i$ models Earth's solid inner core, while the unbounded outer region $\mathcal{V}_o$ provides a simple model for the screening of internal geomagnetic fields by an insulating, rocky mantle. The whole system is taken to rotate around the $z$-axis with angular velocity $\boldsymbol{\Omega}=\Omega \vect{e}_z$.

The motion of the conducting fluid in $\mathcal{V}_s$ is governed by the Navier--Stokes equations under the Boussinesq approximation (which is an approximation often used in geodynamo simulations)
\begin{equation}
\begin{gathered}
\frac{\partial \hat{\vect{U}}}{\partial \hat{t}}+\hat{\vect{U}}\cdot\hat{\nabla}\hat{\vect{U}}= -\frac{1}{\rho_0}\hat{\nabla} \hat{P} -2\Omega \vect{e}_z\times\hat{\vect{U}} + \frac{g_0\alpha \hat{r} }{r_0} \hat{T}\vect{e}_r +\nu \hat{\nabla}^2\hat{\vect{U}} + \frac{1}{\mu}(\hat{\nabla}\times \hat{\vect{B}})\times \hat{\vect{B}},\\
\nabla \cdot \hat{\vect{U}}=0,\\
\frac{\partial \hat{\vect{B}}}{\partial \hat{t}} = \hat{\nabla}\times(\hat{\vect{U}}\times\hat{\vect{B}}) + \eta\hat{\nabla}^2\hat{\vect{B}},\\
\hat{\nabla} \cdot \hat{\vect{B}}=0,\\
\frac{\partial \hat{T}}{\partial \hat{t}}+\hat{\vect{U}}\cdot\hat{\nabla} \hat{T}=\kappa \hat{\nabla}^2\hat{T}.
\label{equ:governing}
\end{gathered}
\end{equation}
These dimensional equations describe the evolution of the velocity field $\hat{\vect{U}}$, temperature field $\hat{T}$, magnetic field $\hat{\vect{B{}}}$ and modified pressure $\hat{P}$ (which also accounts for the centrifugal acceleration). Although we use the temperature, we note that one could also formulate equations (\ref{equ:governing}) in terms of the codensity. In writing the equations in dimensional form, denoted with `hats', we have used the kinematic viscosity $\nu$, the magnetic permeability $\mu$, the magnetic diffusivity $\eta$, the thermal expansion coefficient $\alpha$, the gravity at the outer radius $g_o$, and the thermal diffusivity $\kappa$. The outer region $\mathcal{V}_o$ is taken to be an insulator, and so the electrical current vanishes there: $\hat{\vect{J}}=\hat{\nabla}\times\hat{\vect{B}}=0$, so that the magnetic field satisfies a Laplace equation in $\mathcal{V}_o$. Whilst different approaches have been used to model the inner region $\mathcal{V}_i$, such as taking it to be finitely conducting \citep{Glatzmaier_1995}, we will take it to be insulating.

These equations are non-dimensionalised as follows: length is scaled with the shell width $d=\hat{r}_o-\hat{r}_i$, time with the viscous diffusion time $d^2/\nu$, temperature with the temperature difference $\Delta \hat{T}$ across the shell, and the magnetic field using $\sqrt{\rho \mu \eta\Omega}$. This results in the non-dimensional equations
\begin{equation}
    \begin{gathered}
\textit{Ek}\left( \frac{\partial \vect{U}}{\partial t} +\vect{U}\cdot\nabla\vect{U} -\nabla^2\vect{U}    \right) + 2 \vect{e}_z\times\vect{U} + \nabla P=\tilde{\textit{Ra}}\frac{\vect{r}}{r_0}T+\frac{1}{\textit{Pm}}(\nabla\times \vect{B})\times \vect{B},\\
\nabla\cdot \vect{U} = 0, \\
\frac{\partial \vect{B}}{\partial t} =\nabla\times(\vect{U}\times\vect{B})+\frac{1}{\textit{Pm}}\nabla^2\vect{B},\\
\nabla\cdot \vect{B} = 0,\\
\frac{\partial T}{\partial t}+\vect{U}\cdot\nabla T =\frac{1}{\textit{Pr}}\nabla^2 T,
\end{gathered}
\label{equ:directEquations}
\end{equation}
where we have introduced the notation $\vect{r}=r\vect{e}_r$. 
These equations are governed by the Ekman number $\textit{Ek}$, the modified Rayleigh number ${\tilde{\textit{Ra}}}$ (which accounts for the stabilising effect of rotation on convection, and is related to the classical Rayleigh number $\textit{Ra}=\alpha g_o \Delta T d^3/(\nu \kappa)$ via $\tilde{\textit{Ra}}=\textit{Ra}\textit{Ek}/\textit{Pr}$), the magnetic Prandtl number $\textit{Pm}$ and the Prandtl number $\textit{Pr}$, respectively defined as
\begin{equation}
\textit{Ek}=\frac{\nu}{\Omega d^2}, \qquad \tilde{\textit{Ra}}=\frac{\alpha g_o\Delta \hat{T} d}{\nu\Omega}, \qquad \textit{Pm}=\frac{\nu}{\eta}, \qquad \textit{Pr}=\frac{\nu}{\kappa}.
\label{equ:param}
\end{equation}
The boundary conditions for the temperature and velocity fields are 
\begin{eqnarray}
    T(r_i)&=&1, \\
    T(r_o)&=&0, \\
    \vect{U}(r_i)&=&\vect{0}, \\
    \vect{U}(r_o)&=&\vect{0}.
\end{eqnarray}
We enforce the continuity of the magnetic field through the boundaries with the insulating regions by matching with a potential field at the inner and outer radius, as detailed in \S \ref{sec:numerics}. Under this non-dimensionalisation, the non-dimensional kinetic and magnetic energies take the form
\begin{equation}
K=\frac{1}{2}\iiint \vect{U}\cdot\vect{U}\;\textrm{d}V, \qquad M=\frac{1}{2\textit{Ek}\textit{Pm}}\iiint \vect{B}\cdot\vect{B}\;\textrm{d}V,
\end{equation}
respectively.

\subsection{The adjoint-based optimal control procedure}\label{sec:opt}
We briefly outline the optimisation procedure for determining the initial condition $\vect{q_0}$ that maximises a chosen cost functional under the strong constraint that the system, described by the state variable $\vect{q}(\vect{x},t)$, evolves according to the governing equations, written here in general form:
\begin{equation}
    \matr{M}\frac{\partial \vect{q}}{\partial t}=\boldsymbol{\mathcal{N}}(\vect{q}), \qquad \vect{q}(\vect{x},0)=\vect{q_0}.
    \label{equ:constraint}
\end{equation}
This constrained optimisation problem can be solved by determining the critical points of an augmented Lagrangian
\begin{equation}
\mathcal{L}(\vect{q},\vect{q_0},\vect{q}^\dagger,\vect{q}^\dagger_0)=\mathcal{J}(\vect{q},\vect{q_0})-\left[\vect{q}^\dagger,\matr{M}
    \frac{\partial \vect{q}}{\partial t}-\boldsymbol{\mathcal{N}}(\vect{q})\right]-\langle \vect{q}^\dagger_0,\vect{q}(\vect{x},0)-\vect{q_0}\rangle,
    \label{equ:Lagrangian}
\end{equation}
where $\mathcal{J}$ is the cost functional we need to maximise, $\vect{q}^\dagger$ is a Lagrange multiplier (denoted as adjoint variable), and the inner products $[\cdot,\cdot]$ and $\langle \cdot,\cdot \rangle$ are defined as
\begin{equation}
    [\vect{y}^\dagger,\vect{y}]=\int_0^{t_\textrm{opt}} \iiint (\vect{y}^\dagger)^\mathrm{T}\vect{y}\;\textrm{d}V\textrm{d}t = \int_0^{t_\textrm{opt}} \langle \vect{y}^\dagger,\vect{y}\rangle\textrm{d}t.
\end{equation}
In what follows, the cost functional we will consider is expressed as
\begin{equation}
\mathcal{J}
=\int_0^{t_\textrm{opt}}\iiint\mathcal{J}_I(\mathbf{q})\;\textrm{d}V\textrm{d}t,
\end{equation}
i.e. as the time- and volume- integral of the quantity of interest $\mathcal{J}_I$, with $t_\textrm{opt}$ a chosen time horizon for optimisation. Setting all the first variations of the Lagrangian to zero means that (i) $\vect{q}$ must evolve according to (\ref{equ:constraint}) (also called the ``direct problem''), (ii) $\vect{q}^\dagger$ must solve an adjoint problem, which we will specify below for our particular system, and (iii) the initial condition for the adjoint variable
$\vect{q}^\dagger(\vect{x},0)$ must vanish here:
\begin{equation}
    \frac{\partial \mathcal{L}}{\partial \vect{q}_0} = \vect{q^\dagger_0}=\matr{M}^\mathrm{T}\vect{q}^\dagger(\vect{x},0)=0.
    \label{equ:updateGen}
\end{equation}
In practice, an optimum is iteratively computed by evolving the direct problem forward in time, starting from some initial (typically random) guess $\vect{q_0}$, then evolving the adjoint problem backward in time, down to $t=0$. At this stage, (\ref{equ:updateGen}) provides the required gradient information to update $\vect{q_0}$ in order to increase the value of the cost functional $\mathcal{J}$, and hence move towards a local maximum.

With our direct problem defined by the governing MHD equations (\ref{equ:directEquations}), and using the initial magnetic field $\vect{B}_0=\vect{B}(\vect{x},0)$ as the optimisation variable, we obtain the following adjoint problem (the full details of the derivation of the adjoint equations are contained in appendix \ref{sec:adj}):
\begin{equation}
\begin{gathered}
\textit{Ek}\left(-\frac{\partial \vect{u}^\dagger}{\partial t}+\nabla\times(\vect{U}\times\vect{u}^\dagger)+\vect{u}^\dagger\times\boldsymbol{\omega} - \nabla^2 u^\dagger\right) - 2 \vect{e}_z\times\vect{u}^\dagger + \nabla p^\dagger=-T^\dagger\nabla T +\\
\vect{B}\times(\nabla\times\vect{b}^\dagger)+\frac{\partial\mathcal{J}_I}{\partial \vect{U}},\\
\nabla\cdot \vect{u}^\dagger=0,\\
-\frac{\partial \vect{b}^\dagger}{\partial t}=-\nabla\Pi^\dagger+\frac{1}{\textit{Pm}}\nabla\times(\vect{B}\times\vect{u}^\dagger)-\frac{1}{\textit{Pm}}(\nabla\times\vect{B})\times\vect{u}^\dagger - \vect{U}\times(\nabla\times\vect{b}^\dagger)\\-\frac{1}{\textit{Pm}}\nabla\times(\nabla\times\vect{b}^\dagger)+\frac{\partial\mathcal{J}_I}{\partial \vect{B}},\\
\nabla\cdot \vect{b}^\dagger=0,\\
-\frac{\partial T^\dagger}{\partial t}-\nabla\cdot(\vect{U}T^\dagger)=\tilde{\textit{Ra}}\frac{\vect{r}}{r_0}\cdot\vect{u}^\dagger+\frac{1}{\textit{Pr}}\nabla^2T^\dagger+\frac{\partial\mathcal{J}_I}{\partial T},
\end{gathered}
\label{equ:adjointEqus}
\end{equation}
where $\vect{q}^\dagger=(\vect{u}^\dagger, p^\dagger, \vect{b}^\dagger, \Pi^\dagger, T^\dagger)^\mathrm{T}$ is the adjoint state, and where we have introduced the vorticity $\boldsymbol{\omega}=\nabla\times\vect{U}$.
These adjoint equations, solved backward in time, are ``initialised'' with the final time condition (found by cancelling the first variations of the Lagrangian with respect to $\vect{q}$) that $\vect{u}^\dagger(t_\textrm{opt})=\vect{b}^\dagger(t_\textrm{opt})=\vect{0}$ and $T^\dagger(t_\textrm{opt})=0$. The boundary conditions for the adjoint variables are homogeneous Dirichlet boundary conditions for $T^\dagger$ and all component of $\vect{u}^\dagger$ on both $r=r_o$ and $r=r_i$,
along with vacuum boundary conditions on $\vect{b}^\dagger$ and homogeneous Dirichlet boundary conditions on $\Pi^\dagger$ (see \S\ref{sec:adjoint_derivation}). Note that the integration of the adjoint equations requires that the whole direct solution be stored in memory, which can lead to high memory costs. While we did not encounter this issue within the present study, the memory requirements in future studies might become large enough to prevent the storage of the entire forward solution. This difficulty is classically overcome by the use of checkpointing schemes, which alleviate the large memory requirement at the cost of recomputing the forward solution \citep{Griewank_1992}. Let us mention in particular the {\tt checkpoint\_schedules} \textit{Python} library \citep{Dolci2024}, which can easily be used with our developed code and has implementations of many popular optimal checkpointing schemes - including Revolve \citep{Stumm_2009}, disk-revolve \citep{Aupy_2016}, periodic-disk-revolve \citep{Aupy_2017}, two-level \citep{Pringle_2016}, H-Revolve \citep{Herrmann_2020}, and mixed storage checkpointing \citep{Maddison_2024}.

Throughout the optimisation, we will require that the initial magnetic field is constrained such that $\|\vect{B}_0\|^2=M_0$, where $M_0$ is a specified magnetic energy budget. Whilst this constraint can be handled by introducing an extra Lagrange multiplier \citep{pringle_willis_kerswell_2012}, we will constrain it using Riemannian optimisation methods \citep{Absil_2007}, of which rotation-based projections \citep{Douglas1998,foures_caulfield_schmid_2013} are a special case (see the discussion of \citet{skene2022} for example). In this manner, the Euclidean gradient returned by the optimisation procedure 
\begin{equation}
    \frac{\partial \mathcal{L}}{\partial \vect{B}_0 }= \vect{b}^\dagger(0),
    \label{equ:Euclidean_gradient}
\end{equation}
which provides the direction of steepest ascent for $\mathcal{J}$, is projected into the Riemannian gradient on the hypersphere $\|\vect{B}_0\|^2=M_0$ where optimisation is taking place.

The minimal seed that triggers dynamo action is found by performing a series of optimisations with decreasing values of $M_0$. The initial condition for the hydrodynamic variables are defined as an equilibrium solution of the purely hydrodynamic equations. As our time horizon is smaller than the time needed for a dynamo to be established, we perform one long forward run after each optimisation has converged (following \citet{Mannix2022a}), in order to determine whether the system evolves toward a self-sustained magnetic state. The minimal dynamo seed corresponds to the optimal initial condition with smallest $M_0$ that still triggers a dynamo.

\subsection{Nonlinear travelling wave solutions}\label{sec:Newton}
Many nonlinear states of interest in this system are found to take the form of travelling waves, which rotate around the $z$-axis with the non-dimensional drift frequency $\omega$ relative to our already-rotating reference frame. In order to systematically identify these states, whether they be stable or unstable, we will converge them using the Newton-hookstep algorithm \citep{Dennis_1996}. To that end, we need to adopt the rotating reference frame in which the travelling waves correspond to stationary states: we thus introduce the flow map $\boldsymbol{\Phi}_t(\vect{q}_0,\omega)$ which returns the state at time $t$ starting from initial condition $\vect{q}_0$, in a rotating reference frame with angular frequency $1+\omega$. A travelling wave solution is then an initial condition $\vect{q}_0$ and a drift frequency $\omega$ such that $\boldsymbol{\Phi}_t(\vect{q}_0,\omega)=\vect{q}_0$ for all times $t$, which the Newton-hookstep algorithm identifies by computing the zeros of $\vect{f}(\vect{q}_0,\omega)=\boldsymbol{\Phi}_{t_f}(\vect{q}_0,\omega)-\vect{q}_0$ where the time is now fixed at a small value $t_f$. Although the value of $t_f$ is arbitrary for a travelling wave (as it is stationary in the identified reference frame and so $\boldsymbol{\Phi}_{t_f}(\vect{q}_0,\omega)=\vect{q}_0$ for all $t_f$), a small value is used following the advice of \citet{Willis_NH}, which prevents the algorithm converging to a time-periodic solution.

The classic Newton method finds a root of a vector function $\vect{g}(\vect{x})$, through the iterative procedure,
\begin{gather}
    \vect{x}_{i+1}=\vect{x}_i+\boldsymbol{\delta}\vect{x}_i,\\
    \frac{\partial \vect{g}}{\partial \vect{x}}\boldsymbol{\delta}\vect{x}_i = -\vect{g}(\vect{x}_i).\label{equ:Newton}
\end{gather}
The Newton-hookstep variant aims to alleviate issues that arise when the initial guess $\vect{x}_0$ is far from the true solution, which can cause the algorithm to not converge. This is particularly pertinent for unstable travelling wave solutions where finding a good initial guess is difficult. In the Newton-hookstep algorithm equation (\ref{equ:Newton}) is replaced with
\begin{equation}
    \boldsymbol{\delta}\vect{x}_{i+1} = \underset{
\|\boldsymbol{\delta}\vect{x}\|<\delta}{\textrm{arg min}}\; \left\|\frac{\partial \vect{g}}{\partial \vect{x}}\boldsymbol{\delta}\vect{x} + \vect{g}(\vect{x}_i) \right\|,\label{equ:hookstep}
\end{equation}
where $\delta$ is an (adjustable) trust region size. In this manner, the exact Newton step given by (\ref{equ:Newton}) is relaxed in order to constrain the size of the update to lie within the trust region. The result is a more stable algorithm that has become the primary algorithm of choice for converging unstable equilibria, travelling waves, and relative periodic orbits \citep{VISWANATH_2007,duguet2008,Viswanath_2009,Chandler_Kerswell_2013,Budanur_2017}.

For our system $\vect{x}=(\vect{q}_0,\omega)^\mathrm{T}$, which means that we have one more degree of freedom, given by $\omega$, than equations $\vect{f}$. Therefore, in order to obtain a square system to solve in (\ref{equ:hookstep}) we introduce the additional constraint that the update $\boldsymbol{\delta}\vect{x}_i$ cannot simply rotate $\vect{x}_i$ around the $z$-axis. Rotating $\vect{x}_i(\theta,\phi,r)$ by a small amount around $z$ axis can be achieved by setting $\vect{x}_i(\theta,\phi+\delta\phi,r)$, i.e. we increment $\phi$ by the small amount $\delta\phi$. Using Taylor expansions, one can write
\begin{equation}
\vect{x}_i(\theta,\phi+\delta\phi,r)=\vect{x}_i(\theta,\phi,r) +\nabla\vect{x}_i\cdot \vect{e}_\phi \delta\phi + \mathcal{O}((\delta\phi)^2),
\end{equation}
where $\vect{e}_\phi$ is the unit vector in the $\phi$ direction. This shows that small rotations are given by the term $\nabla\vect{x}_i\cdot \vect{e}_\phi$, and our constraint is that $\boldsymbol{\delta}\vect{x}_i$ should be orthogonal to this direction, which provides the extra equation required for solvability, addressing the underlying rotational symmetry of the underlying system.

\subsection{Numerical implementation}\label{sec:numerics}

All equations are solved using the open-source PDE solver {\tt Dedalus} \citep{Burns2020}. {\tt Dedalus} is a pseudo-spectral solver which is able to solve equations in spherical geometries \citep{Vasil2019,Lecoanet_2019}. {\tt Dedalus} parses the governing equations that have been input by the user in plain text, and provides access to the spectral discretisation and routines for timestepping them. For timestepping we use the second order semi-implicit BDF implicit-explicit multistep scheme \citep{Wang_2008}. Diffusion and pressure-like terms, which allow for divergence-free conditions to be enforced, are timestepped implicitly. In this manner, divergence free conditions are solved for along with all problem variables and do not necessitate any operator splitting methods or poloidal-toroidal decompositions. Spin-weighted spherical harmonics are used for the bases in the $\theta$ and $\phi$ directions, and Jacobi polynomials are used to discretise the radial direction. The typical resolution used in our study consists of all spherical harmonics up to a maximum degree of $\ell_\textrm{max}=63$, Jacobi polynomial degree of $N_\textrm{max}=63$, and a fixed timestep of $\Delta t=0.5\times10^{-4}$. We have verified that our code with this resolution satisfies the Boussinesq hydrodynamic and dynamo benchmark cases of \citet{Christensen2001}.

When solving the direct equations (\ref{equ:directEquations}) we must consider how to handle the divergence free constraint as well as the potential boundary conditions on the magnetic field. In order to enforce the divergence-free condition on $\vect{B}$, we solve for the vector potential, i.e. we solve for $\vect{A}$ such that $\nabla\times\vect{A}=\vect{B}$. The vector potential satisfies the equations
\begin{equation}
    \frac{\partial \vect{A}}{\partial t} =\vect{U}\times\vect{B}+\nabla \phi + \frac{1}{\textit{Pm}}\nabla^2\vect{A},
    \label{equ:vec_potent}
\end{equation}
where $\phi$ is a scalar that is determined here by enforcing the Coulomb gauge condition $\nabla\cdot\vect{A}=0$ although other gauges may be possible \citep{CBT_2020}. Then the potential boundary conditions for the magnetic field on the inner and outer spheres can also be enforced directly on the vector potential \citep{Lecoanet_2019}, taking the form
\begin{eqnarray}
    \left . \left(\frac{\partial \vect{A}_\ell}{\partial r} - \frac{\ell+\sigma}{r}\vect{A}_l \right)\right|_{r=r_i} &=& 0,\\
    \left . \left(\frac{\partial \vect{A}_\ell}{\partial r} + \frac{\ell+1+\sigma}{r}\vect{A}_l \right)\right|_{r=r_o} &=& 0,
\end{eqnarray}
where $\vect{A}_\ell$ is the coefficient of the $\ell^\textrm{th}$ spherical harmonic, and the regularity $\sigma=-1,1,0$ for the $\phi$, $\theta$, and $r$ components.

The vector potential formulation (\ref{equ:vec_potent}) of the direct induction equations is possible since the direct equations naturally preserve the divergence of $\vect{B}$. However, the adjoint equations (\ref{equ:adjointEqus}) do not naturally satisfy this property, with the divergence of the adjoint magnetic field $\vect{b}^\dagger$ being a gauge choice which is enforced with the Lagrange multiplier $\Pi^\dagger$ (see \S \ref{sec:adj} for more details). Hence, we solve directly for $\vect{b}^\dagger$. Note that $\vect{b}^\dagger$ must also satisfy a divergence-free constraint and continuous matching with a potential field on the spherical boundaries. A commonly used, alternative approach \citep[for example]{Dormy_1997,Wicht_2002,Schaeffer_2017} to enforce both conditions at the same time is to decompose the considered field $\vect{X}$ into its poloidal-toroidal form as
\begin{equation}
    \vect{X} = \nabla \times (\mathcal{T}\vect{r}) + \nabla \times (\nabla \times (\mathcal{P}\vect{r})).
\end{equation}
Under this decomposition, the potential boundary conditions take a very simple form in terms of the spherical harmonic decomposition of the poloidal and toroidal parts (see \citet{Hollerbach_2000} for more details).
However as we do not solve the adjoint equations for the poloidal and toroidal parts since it is more natural to solve the full vector form of the equations in {\tt Dedalus}, care is needed in order to properly enforce the potential boundary conditions on $\vect{b}^\dagger$. To that end, we make use of the relations
$\nabla_\parallel^2 (r \mathcal{P}^\dagger) = \vect{e}_r\cdot \vect{b}^\dagger$, and $\nabla_\parallel^2 (r \mathcal{T}^\dagger) =\vect{e}_r\cdot \vect{J}^\dagger$, where $\nabla_\parallel$ is the surface Laplacian, $\vect{J}^\dagger=\nabla\times\vect{b}^\dagger$, and $\mathcal{P}^\dagger$ and $\mathcal{T}^\dagger$ are the poloidal and toroidal parts of the adjoint magnetic field, respectively. By also using the fact that the action of the surface Laplacian on spherical harmonic $Y_\ell^m$ is $\nabla^2_\parallel Y^m_\ell = -[\ell(\ell+1)/r^2 ]Y^m_\ell$ we obtain the following boundary conditions for the adjoint magnetic field:
\begin{eqnarray}
    \left . \vect{e}_r\cdot \left(\frac{\partial \vect{b^\dagger}_\ell}{\partial r} - \frac{\ell-1}{r}\vect{b^\dagger}_\ell \right) \right|_{r=r_i} &=& 0,\\
    \left . \vect{e}_r\cdot \left(\frac{\partial \vect{b^\dagger}_\ell}{\partial r} + \frac{\ell+2}{r}\vect{b^\dagger}_\ell \right)\right|_{r=r_o} &=& 0,\\
    \left . \vect{e}_r\cdot\vect{J^\dagger} \right|_{r=r_o,r_i} &=& 0,
\end{eqnarray}
which indirectly gives the correct conditions on the poloidal and toroidal parts of $\vect{b}^\dagger$. This gives two conditions at each boundary, which, together with the boundary conditions imposed on $\Pi^\dagger$, gives the total number of boundary conditions needed for the adjoint equations.

The numerical implementation of the optimisation procedure outlined in \S \ref{sec:opt} requires a few considerations. Solving the direct equations followed by the adjoint equations provides both the value of the cost functional, and its gradient with respect to the initial magnetic field. This update should not change the specified energy of the initial magnetic field - which, as previously indicated, is handled by directly optimising on a Riemannian manifold. We use for this the {\tt SphereManOpt} library \citep{Mannix_2024}, similar to {\tt Pymanopt} \citep{Townsend_2016} but focusing directly on spherical manifolds, with a conjugate gradient algorithm from \citep{Sato_2022} and Armijo line search \citep{Wright_1999}.

For implementing the Newton-hookstep algorithm we use the routines available in the JFNK-Hookstep Github repository \citep{Willis_2019,Willis_NH}. Specifically, we use the FORTRAN implementation of the code developed for Openpipeflow \citep{Willis_2017}, which is compiled into a Python module using {\tt f2py} \citep{Peterson_2009}. This enables the algorithm to be easily used together with {\tt Dedalus}. The flow map $\boldsymbol{\Phi}_{t_f}(\vect{q}_0,\omega)$ is provided to the Newton-hookstep algorithm by using {\tt Dedalus} to solve equations (\ref{equ:directEquations}) with the Coriolis term changed to $(2+2\omega)\vect{e}_z\times\vect{U}$, and the boundary conditions for $\vect{U}$ now taking the form $\vect{U}(r=r_i)=-\textit{Ek}^{-1}\omega r_i\sin\theta \vect{e}_\phi$ and $\vect{U}(r=r_o)=-\textit{Ek}^{-1}\omega r_o\sin\theta \vect{e}_\phi$. When the Newton--hookstep algorithm converges, this corresponds to solving the equations in a reference frame that rotates with the rotation rate of the travelling wave, rather than the Earth. Hence, the fluid will appear stationary, with the boundary conditions giving the rotation rate of Earth relative to the travelling wave.  Note that as with the direct equations (\ref{equ:directEquations}), the centrifugal term is absorbed in the pressure gradient. While there are sophisticated ways for generating initial guesses for the Newton-hookstep algorithm, including those based on near recurrences \citep{VISWANATH_2007,Chandler_Kerswell_2013}, dynamic mode decomposition \citep{Page_Kerswell_2020}, variational methods \citep{Lan_2004, Parker_Schneider_2022}, and convolutional autoencoders \citep{Page_2021, Page_Holey_Brenner_Kerswell_2024}, at the parameters used for this study we can find initial guesses directly from snapshots obtained from the evolution of the governing equations. The travelling waves are converged to a tolerance of $10^{-6}$, with error measured using the $L_2$ norm of the spectral discretisation of the fields. In order to calculate the stability of each of the travelling wave solutions, we take a matrix-free approach similar to that described by \citet{skene_tobias_2023}. The stability problem is solved in a reference frame corotating with the travelling wave, reducing a Floquet stability problem to that of a fixed point.
\section{Results}\label{sec:results}

\subsection{Benchmark setup}

For the purpose of the present study, and to test the numerical scheme, we adopt the parameters of the dynamo benchmark introduced by \citet{Christensen2001}: $\textit{Ek}=10^{-3}$, $\tilde{\textit{Ra}}=100$,  $\textit{Pm}=5$, $\textit{Pr}=1$ and $r_i/r_o=0.35$. This benchmark is a widely studied subcritical dynamo solution, and therefore provides an ideal case on which to test our optimisation procedure. The initial temperature field for the benchmark is specified as
\begin{equation}
    T=\frac{r_ir_o}{r}-r_i + \left( a(1-3x^2+3x^4-x^6)Y_4^4(\phi,\theta) + \textrm{c.c} \right),
    \label{equ:therm_ic}
\end{equation}
with $x=2r-r_i-r_o$, and $a=0.1$, driving an $m=4$ convective instability (`case 0' of \citet{Christensen2001}: non-magnetic convection). While the resulting hydrodynamic state is linearly stable to magnetic perturbations, it is (nonlinearly) unstable to some finite-amplitude magnetic perturbations that trigger dynamo action. \citet{Christensen2001} suggest that the basin of attraction of this dynamo solution may be small and therefore recommend prescribing a mainly dipolar, initial magnetic field of the form
\begin{eqnarray}
    \mathcal{T} &=& \frac{4\sqrt{5\pi}}{3}\sin(\pi[r-r_i])Y_2^0(\phi,\theta), \label{equ:base_toroidal}
\\
    \mathcal{P} &=& -\frac{5}{4r^2}\sqrt{\frac{\pi}{3}}(r^3[3r-4r_o]+r_i^4)Y_1^0(\phi,\theta),\label{equ:base_poloidal}
\end{eqnarray}
with (dimensionless) magnetic energy $M_0=1215$, which kickstarts an $m=4$ dynamo solution. We note that, as at the benchmark parameters the system is hydrodynamically unstable to multiple modes of convection, other dynamo solutions are possible using different thermal forcing wavenumbers --- see \citet{Feudel2017} for instance. In fact, we have checked that our procedure works without modification for identifying solutions with $m=5$ symmetries by replacing the $Y_4^4$ spherical harmonic in (\ref{equ:therm_ic}) with $Y_5^5$. However, as the overall dynamical picture is identical to the $m=4$ case, we do not present it here.

\begin{table}
\centering
\begin{tabular}{c c c c c c} 
 Label & $\textit{K}$ & $\textit{M}$ & $\omega$ & Stable & Growth rate\\
 \hline
TW0 & $58.35$  & $0$     & $0.000182$ & Yes & -2.06 (-0.80)\\
TW1 & $30.773$ & $626.4$ & $-0.00310$ & Yes & -0.56\\
TW2 & $29.72$  & $459.9$ & $-0.00250$ & No  & 1.13\\
 \hline
\end{tabular}
\caption{Travelling wave solutions to the governing equations with $m=4$ symmetry. $\omega$ is the frequency at which they drift relative to the reference frame. $K$ and $M$ give their dimensionless kinetic and magnetic energies, respectively. Also displayed are the (dimensionless) least stable growth rates of the solutions to perturbations. For TW0, the growth rate is displayed for a purely hydrodynamic perturbation, with the growth rate for a purely magnetic perturbation also shown in brackets. With our definition of $\omega$, the numerical values differ from the drift frequencies of \citet{Christensen2001} by a factor of $\textit{Ek}$.}
\label{table:travelling_waves}
\end{table}

\begin{figure}
  \centering
  \begin{tabular}{ccc}
  \begin{subfigure}[t]{0.29\textwidth}
      \includegraphics[height=4cm]{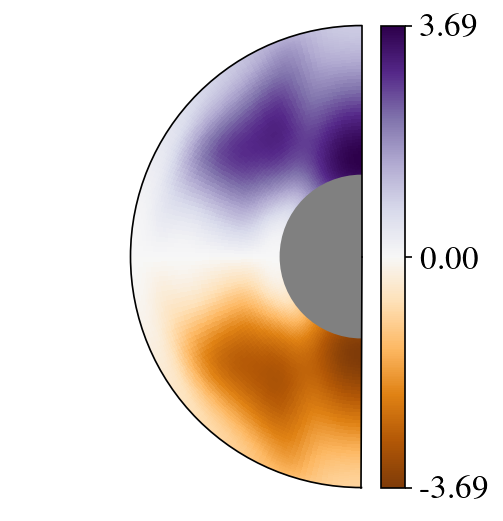}
      \caption{Meridional slice of the radial component of the magnetic field.}
    \end{subfigure}
    &
    \begin{subfigure}[t]{0.29\textwidth}
      \includegraphics[height=4cm]{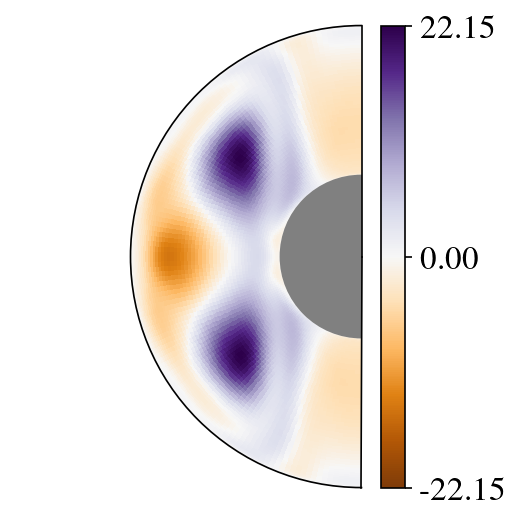}
      \caption{Meridional slice of the radial component of the current.}
    \end{subfigure}
  &
    \begin{subfigure}[t]{0.39\textwidth}
      \includegraphics[height=4cm]{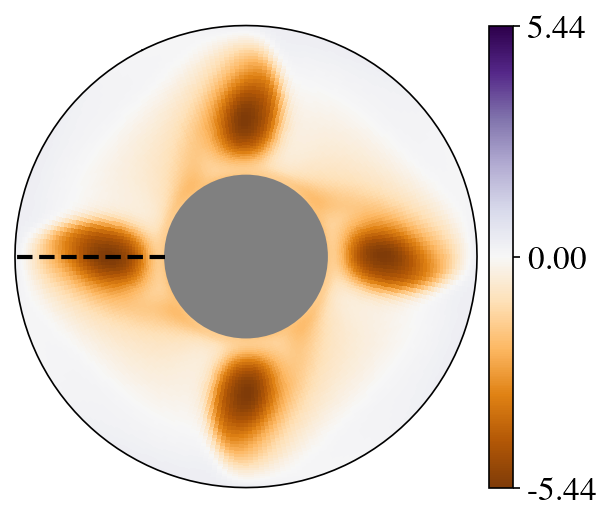}
      \caption{Equatorial slice of the latitudinal component of the magnetic field.}
    \end{subfigure}
  \end{tabular}
  \caption{Slices of the $m=4$ nonlinear dynamo state (TW1). The dashed line on the equatorial slice indicates where the meridional slices were taken.
  }
  \label{fig:stable_dynamo}
\end{figure}

Using the Newton-hookstep procedure outlined in \S \ref{sec:Newton}, we find the system to possess three $m=4$ travelling wave (TW) solutions at our working parameters, characterised by the energies and drift frequency (relative to the already-rotating reference frame) displayed in table \ref{table:travelling_waves}. The table shows one purely hydrodynamic solution (TW0) which corresponds to the `case 0' benchmark of \citet{Christensen2001}. This hydrodynamic solution arises via a supercritical Hopf bifurcation at $\tilde{\textit{Ra}}_c=55.9$ (see the weakly nonlinear analysis of \citet{skene2024}, for instance), and is linearly stable to both hydrodynamic and magnetic perturbations at our working parameters (with growth rates displayed in table \ref{table:travelling_waves}). The two other TW solutions are MHD solutions that form at a saddle node bifurcation at $\tilde{\textit{Ra}}\sim 98$ \citep{Feudel2017} connected to the purely convecting solution: one stable (TW1) and corresponding to the dynamo  (`case 1' of \citet{Christensen2001}); the other unstable (TW2). Figure \ref{fig:stable_dynamo} shows slices of TW1, showing that the nonlinear state is mainly dipolar, with a quadrupolar toroidal part (evident from the radial component of the current). We note here that TW2 has a very similar structure to TW1, which is to be expected since we are close to the saddle node bifurcation, although it shows slightly more spatial localisation than TW1. This localisation becomes more apparent at higher Rayleigh numbers, where the two states start to deviate in structure.

\begin{figure}
	\centering
	\includegraphics[]{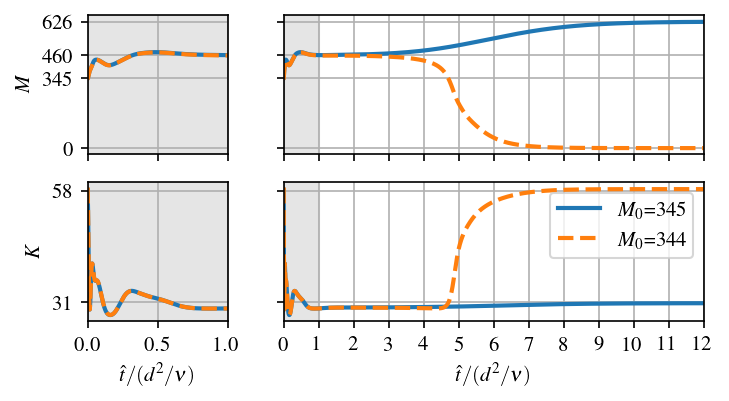}
	\caption{Magnetic energy (top) and kinetic energy (bottom) evolution with the rescaled benchmark with two different initial energies.}
    \label{fig:baseline}
\end{figure}

We now let the system evolve to the saturated, purely hydrodynamic, $m=4$ convective solution (TW0). This state is then seeded with a magnetic field of the form given by \citet{Christensen2001} (\ref{equ:base_toroidal})-(\ref{equ:base_poloidal}); however, the field amplitude is rescaled to have a given initial magnetic energy $M_0$, which is gradually decreased (through a series of repeated simulations) down to the point where the initial field fails to trigger a dynamo. The energy timeseries for two slightly different energy budgets close to the threshold are shown in figure \ref{fig:baseline}. Initially, the evolution is very similar for both cases and is characterised by a transient amplification of the magnetic field, which extracts energy from the convection. This transient growth occurs on a dynamical time scale and is followed by a plateau until around $t \approx 2.5$ (i.e. half an ohmic timescale). At this point the evolutions differ, with the $M_0=345$ case achieving further growth in magnetic energy to reach a new stable MHD state (corresponding to the TW1 dynamo) over a few ohmic timescales, and the $M_0=344$ case magnetic field decaying back to the purely hydrodynamic state TW0 (for future reference we label the rescaled benchmark initial condition with $M_0=344$ RB). Remarkably, the flow state around $t \approx 2.5$ corresponds to the unstable MHD state TW2. By performing a stability calculation of TW2, we confirm that it has a single unstable direction (see table \ref{table:travelling_waves}). Thus, TW2 is found to be a saddle point with both solutions initially approaching TW2 along its stable directions. The flow can then either approach TW0 or TW1 along this unstable direction, with which travelling wave the flow approaches depending on which side of the saddle the state lies.

\subsection{Optimised seeds}
With the baseline behaviour for our system established, we now seek to identify the spatial structure of the lowest-energy magnetic perturbation required to reach the dynamo solution, using the optimisation procedure outlined in $\S$\ref{sec:opt}. In other words, what is the smallest magnetic field initial condition, with the hydrodynamic initial condition fixed as TW0, that leads to TW1? Following \citet{Mannix2022a}, we initially consider to that end the cost functional with cost integrand
\begin{equation}
    \mathcal{J}_I = \frac{1}{2\textit{Ek}\textit{Pm}}\vect{B}\cdot\vect{B},
    \label{equ:JI_int}
\end{equation}
which corresponds to maximising the time-integrated magnetic energy.  As our expected magnetic field is large scale, we initialise the optimisation procedure with a random guess with spherical harmonic degrees up to $\ell=m=2$, and a random radial dependence with Jacobi polynomial degree up to four. We note here, that we have tested random initial guesses up to $\ell=m=4$. The behaviour of the optimisation is initially similar, with the seed magnetic field converging to a large scale solution. However, the less sparse spectra of the $\ell=m=4$ guess causes slower convergence, with the optimisation spending a long time removing the increased amount of fine scale components which build up quicker when using this guess. Even though this guess is still large-scale, the initial strongly nonlinear transient period, in which the magnetic field takes energy from the convection, leads to updates which initially increase the higher-scale components of $\vect{B}_0$. We further describe the convergence behaviour of the optimisation routine in appendix \ref{sec:convergence}. Based on the quick evolution of transients in the previous section, we start by considering a short optimisation window of $t_\textit{opt}=0.2$. The initial magnetic energy budget is specified to be $M_0=344$ as this is below the threshold at which a dynamo can be reached from the baseline solution. 

\begin{figure}
	\centering
	\includegraphics[]{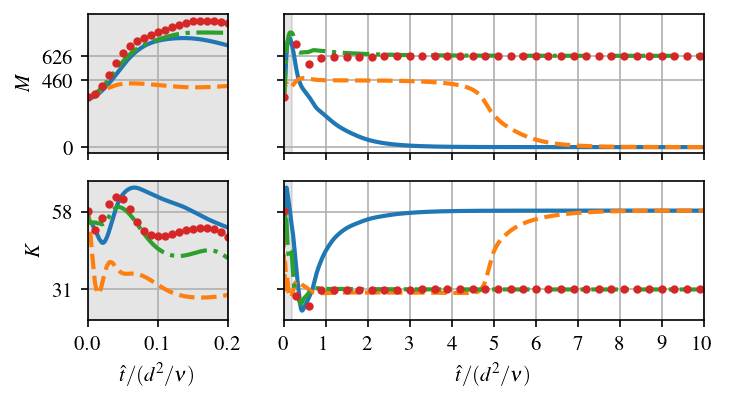}
	\caption{Comparison of the magnetic energy (top) and kinetic energy (bottom) evolution with four different initial conditions with $M_0=344$. The initial conditions are the optimised seeds obtained by optimising the total magnetic energy with a random initial guess (solid blue line), optimising the total magnetic energy with an initial guess of the rescaled benchmark initial condition (RB) (dotted red line), and optimised seed obtained by optimising the energy in the $m=0$ part of the magnetic field starting from a random guess (dashed-dotted green line). We also show the timeseries obtained without optimisation, starting directly from the rescaled benchmark initial condition RB (dashed orange line). The plots on the left show the energy evolution in the optimisation window $t_\textrm{opt}=0.2$, and the plots on the right show the long-time evolution.}
    \label{fig:opt_evolution}
\end{figure}

\begin{figure}
  \centering
  \begin{tabular}{ccc}
  \begin{subfigure}[t]{0.29\textwidth}
      \includegraphics[height=4cm]{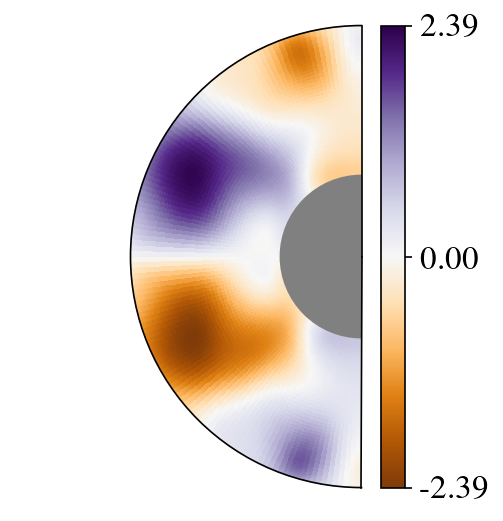}
      \caption{}
    \end{subfigure}
    &
    \begin{subfigure}[t]{0.29\textwidth}
      \includegraphics[height=4cm]{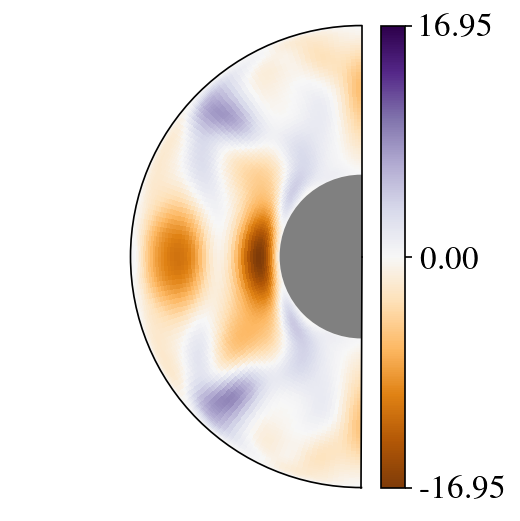}
      \caption{}
    \end{subfigure}
  &
    \begin{subfigure}[t]{0.39\textwidth}
      \includegraphics[height=4cm]{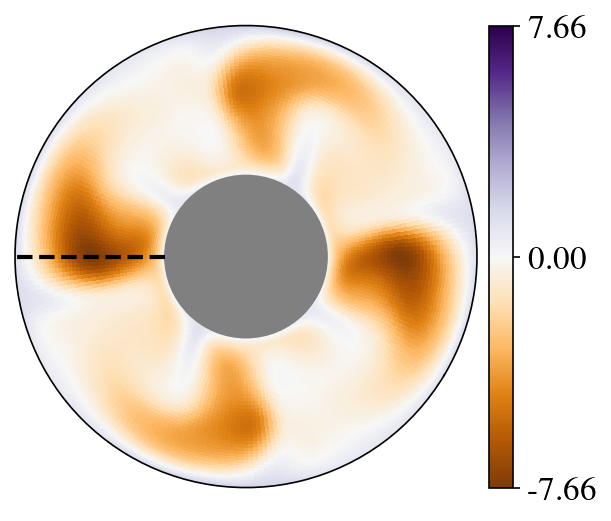}
      \caption{}
    \end{subfigure}\\
  \begin{subfigure}[t]{0.29\textwidth}
      \includegraphics[height=4cm]{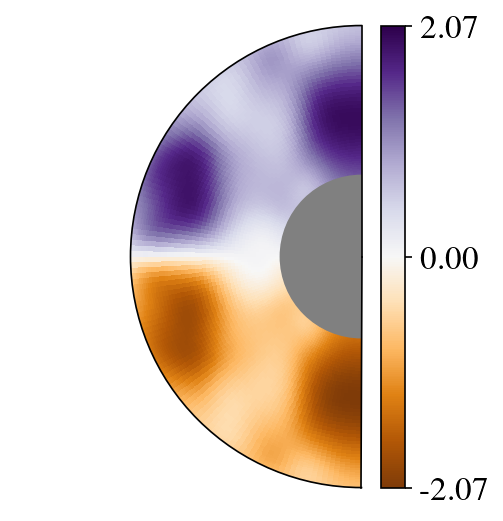}
      \caption{}
    \end{subfigure}
    &
    \begin{subfigure}[t]{0.29\textwidth}
      \includegraphics[height=4cm]{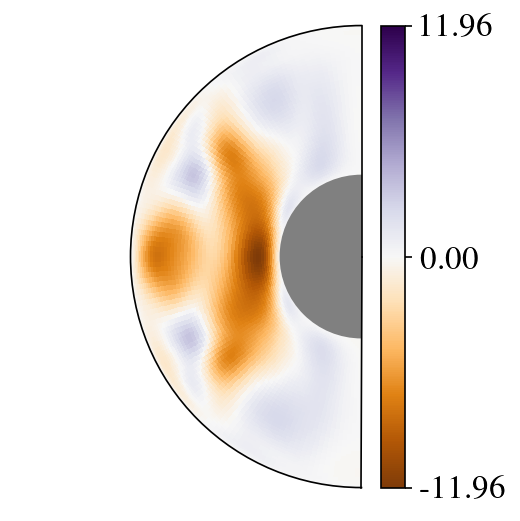}
      \caption{}
    \end{subfigure}
  &
    \begin{subfigure}[t]{0.39\textwidth}
      \includegraphics[height=4cm]{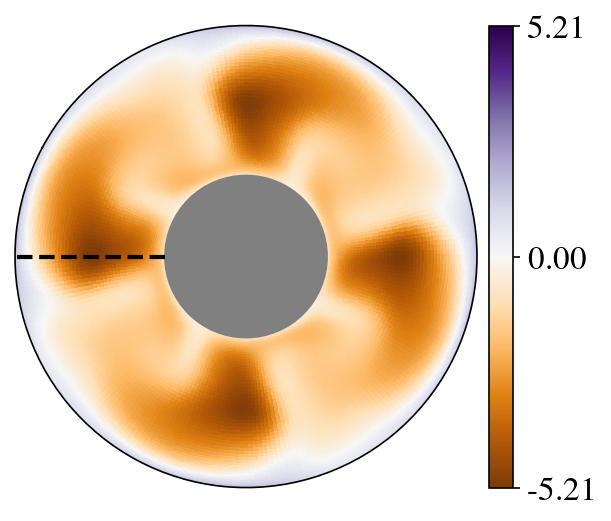}
      \caption{}
    \end{subfigure}\\
    \begin{subfigure}[t]{0.29\textwidth}
      \includegraphics[height=4cm]{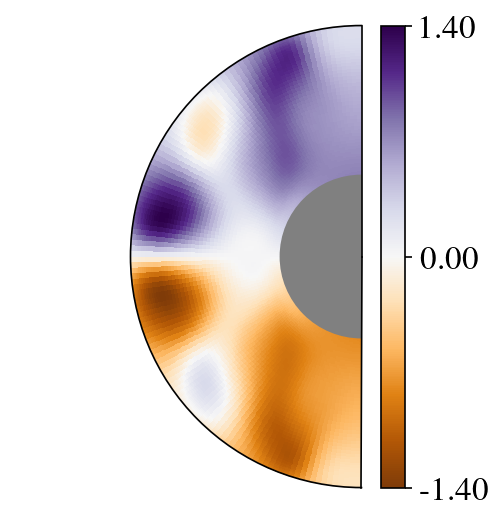}
      \caption{}
    \end{subfigure}
    &
    \begin{subfigure}[t]{0.29\textwidth}
      \includegraphics[height=4cm]{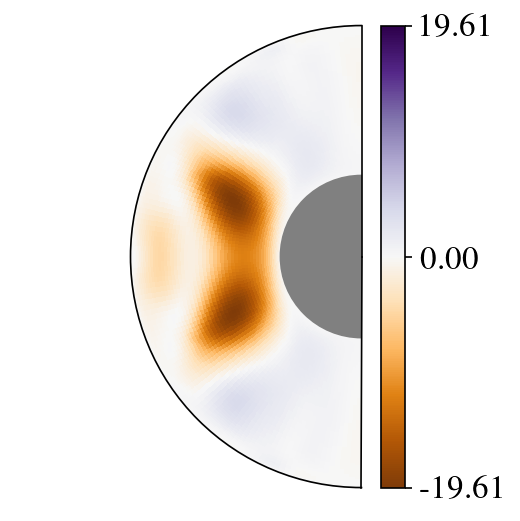}
      \caption{}
    \end{subfigure}
  &
    \begin{subfigure}[t]{0.39\textwidth}
      \includegraphics[height=4cm]{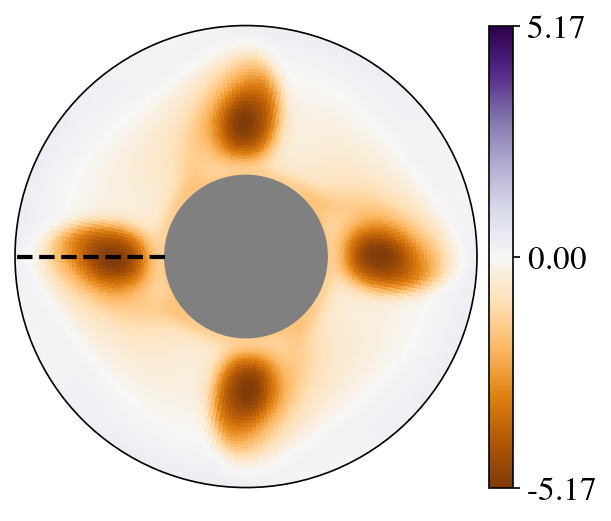}
      \caption{}
    \end{subfigure}
  \end{tabular}
  \caption{\textit{Top:} Slices of an example optimal seed (non-dynamo) obtained with the total magnetic energy cost functional, with $t_\textrm{opt}=0.2$ and $\textit{M}_0=344$. \textit{Middle:} Slices of the optimal seed (dynamo) obtained with the axisymmetric magnetic energy cost functional with $t_\textrm{opt}=0.2$ and $\textit{M}_0=344$. \textit{Bottom:} Slices of the optimal seed (dynamo) obtained with the axisymmetric magnetic energy cost functional with $t_\textrm{opt}=0.4$ and $\textit{M}_0=162$. Figures (a), (d) and (g) show meridional slices the radial component of the magnetic field, figures (b), (e) and (h) show meridional slices of the radial component of the current, and figures (c), (f) and (i) show equatorial slices of the latitudinal component of the magnetic field. The dashed lines on the equatorial slices indicates where the meridional slices were taken.}
  \label{fig:alpha_opt}
\end{figure}

Because the optimisation procedure only identifies local extrema, it was repeated multiple times starting from different realisations of the noisy initial guess. This resulted in identifying several optimised seeds (all well converged), corresponding to distinct local extrema. While all the optimised seeds did trigger strong transient growth of the magnetic energy when used as initial conditions for direct simulations, it was nevertheless found upon long-time integration ($t \gg t_\textrm{opt}$) that most of these seeds did not successfully kickstart a dynamo. 
Figure \ref{fig:opt_evolution} shows (solid, blue line) the magnetic and kinetic energy evolution starting from a (ultimately non dynamo) total-energy-based optimal seed. The figure shows that, despite success of the optimisation leading to the transient growth of magnetic energy being considerably larger than that of the baseline solution (dashed, orange line), the magnetic energy eventually decays and no dynamo solution is reached. We also show (dotted, red line) the magnetic and kinetic energy evolution of an optimised seed that reaches a dynamo solution. This solution shows increased transient growth in the optimisation window and eventually reaches a dynamo solution, showcasing the coexistence of multiple local optima.

Increasing the time horizon of the optimisation procedure up to $t_\textrm{opt}=0.4$ was not found to add robustness to the computation of optimal seeds as multiple local extrema characterised by strong transient growth (but not necessarily a long-time dynamo) continue to coexist at such target times. While it is tempting to further increase $t_\textrm{opt}$ --- indeed we expect that optimising over (very) long times would ultimately discriminate between dynamo seeds and magnetic perturbations yielding only transient growth and ultimately decay -, for long target times convergence becomes increasingly difficult (numerical cost notwithstanding). It is therefore more realistic to define another cost functional that could possibly reduce the number of local extrema, yielding more robust seed identification. This is consistent with the observation of \citet{kerswell2014} that ``the measure used as the objective functional and the norm constraining the initial condition do not need to be the same" for computing minimal seeds.

To that end, we now consider an alternative cost functional and optimise the energy in the $m=0$, or axisymmetric, part of $\vect{B}$. This can be achieved by considering
\begin{equation}
    \mathcal{J}_I = \frac{1}{2\textit{Ek}\textit{Pm}}\bar{\vect{B}}\cdot\bar{\vect{B}},
    \label{equ:JI_m_int}
\end{equation}
with 
\begin{equation}
\bar{\vect{B}} = \frac{1}{2\pi}\int_0^{2\pi} \vect{B} \;\textrm{d}\phi.
\end{equation}
This cost functional has a physical motivation. It is known that dynamos with a strong magnetic field have a tendency to be more dipolar and large scale. Optimising the zonal mean magnetic energy promotes flows that are large scale and on which diffusion operates slowly; this will promote the possibility of dynamo action with a strong field. It is hoped that this physically motivated cost functional will therefore give more robust results in finding the subcritical dynamo.

The result of optimising this cost functional is also shown in figure \ref{fig:opt_evolution} (dashed-dotted, green line), and we label this optimal seed OS1. In contrast to optimising the total magnetic energy, optimising this new cost functional is robust in the sense that it was found to always identify the same optimal seed (regardless of the random initial guess used to initialise the algorithm). Furthermore, the computed optimal seed does indeed lead to a dynamo. Incidentally, it is also found to outperform some ``optimal'' seeds that (locally) maximise the total magnetic energy (as exemplified in figure \ref{fig:opt_evolution}). Based on these observations, one could perhaps speculate that using this new cost functional \textit{might} be a way to `smooth out' the optimisation landscape, possibly suppressing the multiple local extrema and thus providing a good proxy for the global extremum of the first (total energy) optimisation problem --- although this is highly uncertain. 

The optimal `non-dynamo' seed obtained with the total energy cost functional and optimal dynamo seed obtained by maximising the axisymmetric energy are shown in figure \ref{fig:alpha_opt}. Clearly, there are significant differences between the two seeds, with the non-dynamo and dynamo seeds having a dominant quadrupolar and dipolar poloidal part, respectively. From the equatorial slice it is also evident that the non-dynamo seed has strong $m=2$ and $m=4$ components, whereas the dynamo seed is predominantly $m=4$. This visual comparison is also confirmed by computing the spectra of each seed, which also reveals the presence of a strong quadrupolar toroidal field for the dynamo seed (similarly to the benchmark field). Although not shown, the optimal dynamo seed obtained with the total energy cost functional has a similar structure to that obtained using the axisymmetric energy cost functional. From figure \ref{fig:opt_evolution} we see that the non-dynamo seed does not attract to the edge state and quickly decays to TW0, which can now be attributed to it having a different symmetries than that of TW1 and TW2. Although the dynamo seeds found with the total energy and axisymmetric energy cost functionals are similar, we also see from figure \ref{fig:opt_evolution} that the seed obtained by optimising the total energy shows more transient growth in the optimisation window. This is to be expected, as it is found by directly optimising the total energy, whereas the seed obtained with the axisymmetric energy only indirectly maximises this quantity. However, we also see that after the optimisation window the $m=0$ seed more directly approaches the dynamo. This indicates that the slightly differing structure of the total-energy optimising dynamo seed, whilst leading to more transient growth, comes at the expense of not directly targeting a dynamo solution. The result is that the $m=0$ energy based cost functional is the more appropriate choice for finding seeds that lead to dynamo action with short optimisation time horizons.

\begin{figure}
  \centering
  \begin{tabular}{cc}
  \begin{subfigure}[t]{0.45\textwidth}
      \includegraphics[width=\textwidth]{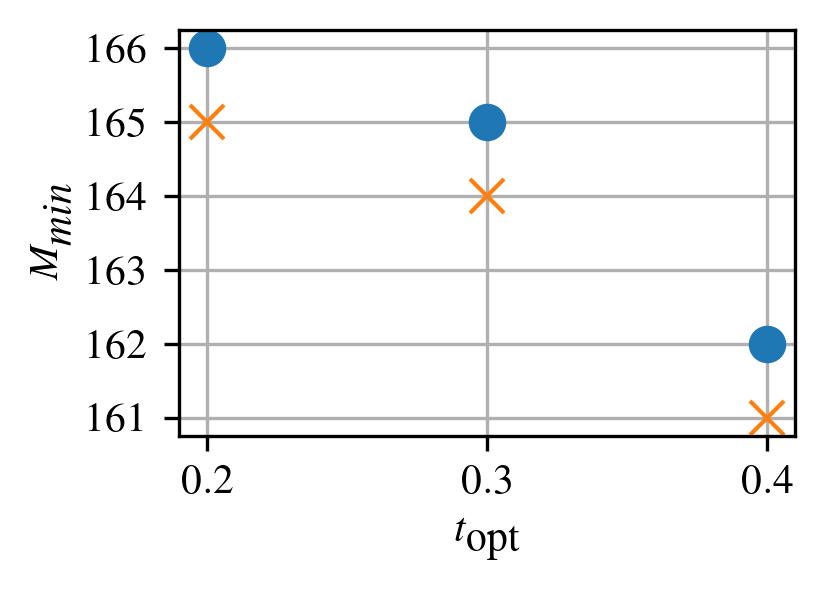}
    \end{subfigure}
    &
    \begin{subfigure}[t]{0.45\textwidth}
      \includegraphics[width=\textwidth]{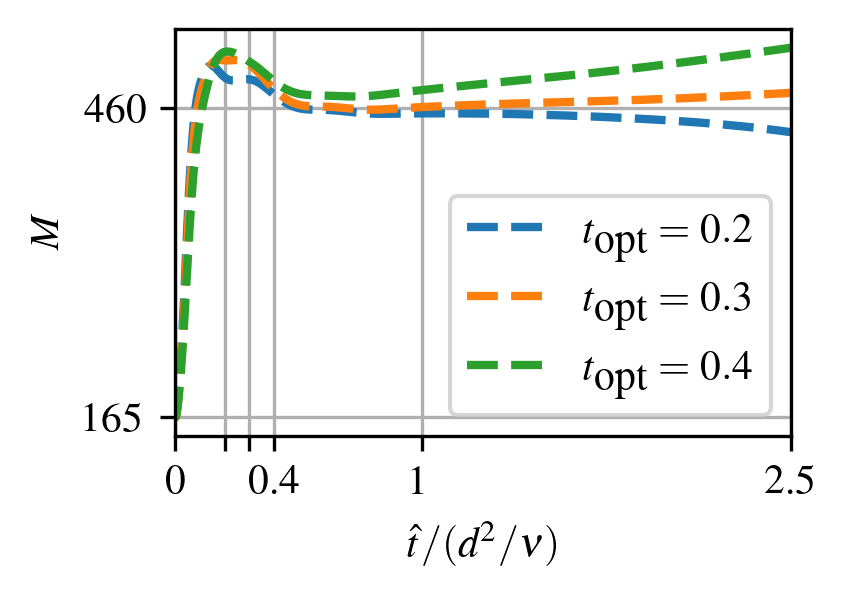}
    \end{subfigure}
  \end{tabular}
  \caption{Robustness of the optimisation results with respect to the time horizon $t_\textrm{opt}$. \textit{Left:}~The minimum initial magnetic energy budget to find a dynamo solution for a given time horizon $t_\textrm{opt}$ ($m=0$ cost functional). Circles show runs that succeed in finding a dynamo, and crosses indicate no dynamo was found. \textit{Right:}~Magnetic energy timeseries for $M_0=165$ starting from optimised seeds ($m=0$ cost functional) computed with different time horizons.}
  \label{fig:dynamovsRT}
\end{figure}

\subsection{`Minimal' seeds and dynamical landscape}

With a robust cost functional identified, we now turn our attention to lowering the energy of the initial seed. To this end, we perform a series of optimisations maximising the total axisymmetric energy, for gradually decreasing initial energy budgets at fixed time horizon. Each optimal seed then serves as initial condition for a long-time DNS: if a dynamo state is found at long times, then $M_0$ is decreased further and a new optimisation is performed. We thus determine the initial energy $M_0^*$ below which the computed optimal seed no longer triggers a subcritical dynamo, and repeat the whole procedure for increasing time horizon. Indeed \citet{Pringle_2010,kerswell2014} already noted that the choice of time horizon is crucial in accurately determining the amplitude of the minimal seed for transition to turbulence. As shown in Figure \ref{fig:dynamovsRT} \textit{(Left)}, $M_0^*$ was found not to vary much as the target time was increased up to $t_\textrm{opt}=0.4$. Moreover, the effect of lengthening the optimisation time horizon is shown in Figure \ref{fig:dynamovsRT} \textit{(Right)} for an initial energy budget of $\textit{M}_0=165$: at this energy $t_\textrm{opt}=0.2$ is not long enough to optimise an initial condition which yields a dynamo. We see that lengthening the time horizon to $t_\textrm{opt}=0.3$ and $t_\textrm{opt}=0.4$ enables the optimisation procedure to converge to a dynamo solution. Although the early timeseries in Figure \ref{fig:dynamovsRT} \textit{(Right)} show that there are differences in the magnetic energy evolution as the $t_\textrm{opt}$ is changed, we see that these changes become significantly less as $t_\textrm{opt}$ is increased. This indicates that the computed extremum eventually becomes insensitive to $t_\textrm{opt}$, which we can attribute to the fact that for $t_\textrm{opt}\approx0.4$ the transient growth regime has ended and the flow has approached the unstable MHD state TW2. For longer time-horizons the optimisation procedure struggles to converge. 

Based on these findings we conclude that $t_\textrm{opt}=0.4$ is the most practical choice of time horizon, enabling the optimisation procedure to converge, while at the same time being long enough to ensure robustness with respect to changes in $t_\textrm{opt}$. Recall, this is non-dimensionalised with respect to the viscous diffusion time, and this is a dynamical timescale. In other words, we only need to optimise up to when the state approaches the unstable edge state, at which point we require the instability of the edge state to lead to the dynamo solution. This is similar to previous studies finding minimal seeds for flows with edge states (see the work of \citet{Duguet_2010, JUNIPER_2011} for example). The spatial structure of the `minimal' seed (i.e. the lowest-energy dynamo seed we could find) is shown in figure
\ref{fig:alpha_opt} \textit{(Bottom)}
; it is similar to the optimal seed found at higher energy (OS1) $\textit{M}_0=344$ (figure \ref{fig:alpha_opt}, \textit{Middle}), however shows more localisation (indicating a more concentrated magnetic energy density profile in all spatial directions). We note that neither of these magnetic fields closely resemble either the nonlinear dynamo state (figure \ref{fig:stable_dynamo}) or unstable state (similar to figure \ref{fig:stable_dynamo}). We see that although they have some similar features, such as a dominant dipolar poloidal part, and quadrupolar toroidal part, the radial structure and energy spectra is significantly different between the two solutions - and is more complex in the case of the optimal seed. This showcases the need for a systematic procedure that can work from random initial guesses \citep{Mannix2022a}. The minimal seed, obtained with $M_0=162$ and $t_\textrm{opt}=0.4$ is labelled OS2. Additionally, we also label the optimal seed obtained with $M_0=161$ and $t_\textrm{opt}=0.4$, which does not reach a dynamo solution, OS3.

\begin{figure}
\includegraphics[width=\textwidth]{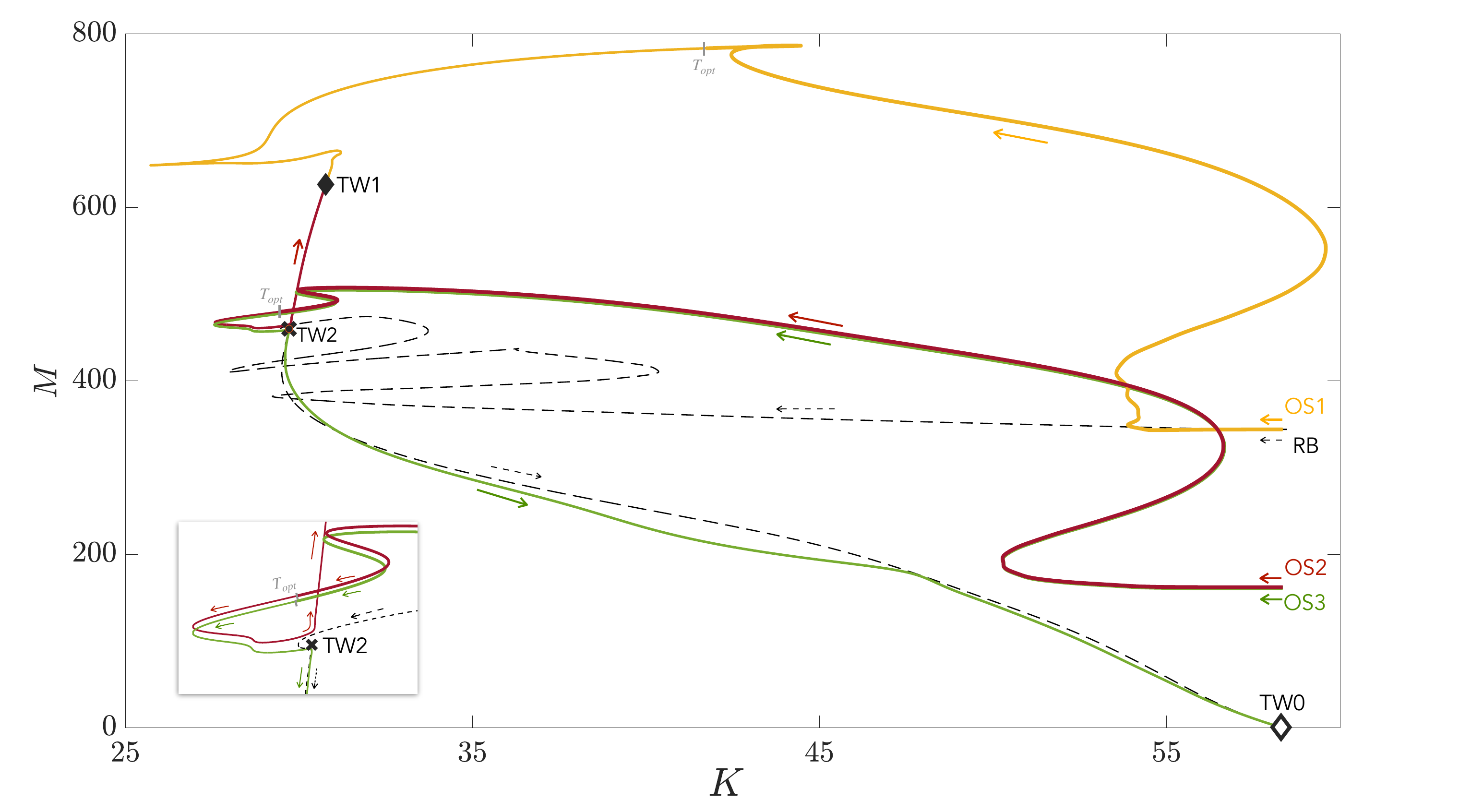}
      \caption{Dynamical landscape, projected in the kinetic energy ($K$) versus magnetic energy ($M$) phase space. The three travelling waves states are signalled by TW0 (empty diamond; linearly stable hydrodynamic state), TW1 (full diamond; linearly stable dynamo state) and TW2 (black cross; linearly unstable MHD state, edge state). Four trajectories are shown, corresponding to simulations initiated with, respectively, a rescaled benchmark magnetic field RB ($M_0=344$, dashed black line), and three optimal seeds identified with the $m=0$ energy cost:  OS1 (solid yellow line; $M_0=344$, $t_\textrm{opt}=0.2$), the minimal dynamo seed OS2 (solid red line; $M_0=162$, $t_\textrm{opt}=0.4$) and OS3 (solid green line; $M_0=161$, $t_\textrm{opt}=0.4$). Thicker lines mark the span of the time horizons for the optimisation procedures.
      }
  \label{fig:landscape}
\end{figure}

\section{Conclusions}\label{sec:conclusions}

Motivated by the search for strong branch solutions of the geodynamo, we have developed the code to perform adjoint-based optimal control on a convection-driven, magnetohydrodynamic flow in a rotating spherical shell, using the {\tt Dedalus} environment. We have revisited the well-studied dynamo benchmark of \cite{Christensen2001} as an example case to systematically identify minimal magnetic perturbations (in the sense of least magnetic energy) that nonlinearly trigger a convective dynamo, relying on transient growth optimisation.

Whereas in the systems considered by \cite{Mannix2022a}, maximising the time-integrated magnetic energy over a fraction of the ohmic timescale was found suitable to robustly identify dynamo solutions and their minimal seeds, in the present system this approach appears less reliable. Indeed, the optimisation procedure identified multiple magnetic conditions that (locally) maximise magnetic energy growth but do not trigger a sustained dynamo at longer times.
We are facing here two practical difficulties: firstly, the optimisation problem is highly nonconvex; secondly, we are restricted to relatively short time horizons for the optimisation. Indeed we found optimisation to become highly sensitive to small changes in the initial conditions as the target time approached $t_\textrm{opt}=0.4$, rendering convergence difficult; this observation is however readily explained by the fact that the system was approaching the (unstable) edge state by this time. Moreover, our use of a differentiate-then-discretise approach to solve the adjoint equations \citep{Mannix_2024} causes our gradient estimate to deteriorate as the time horizon is increased. Yet, the optimisation question we are truly interested in is best investigated over (very) long time horizons: indeed the very definition of dynamo action is the existence of an initial condition such that magnetic energy does not vanish as time goes to infinity! Were $t_\textrm{opt}$ allowed to become long enough while maximising the magnetic energy, then all initial conditions yielding failed dynamos would essentially yield negligible objective functions, while any local maximum would have to be a dynamo seed. Whereas for short target times, maximising the total magnetic energy does not penalise local extrema corresponding to initial conditions that trigger strong transient growth ultimately followed by decay. In other words, we need for practical reasons to devise a cost functional that, when extremised over relatively \textit{short} time horizon, acts as a robust proxy for discriminating \textit{long} times evolution.

This situation is reminiscent of a similar issue encountered in optimal mixing problems, where multiscale cost functions can act as shorter-target time proxies for long-time flow evolution. Specifically, minimising Sobolev norms of negative indexes called the \textit{mix-norms} \citep{Matthew05,DT06} by a short target time has been shown to effectively yield the same results (in terms of both the optimal perturbations and the achieved objective functions) as minimising the flow variance over long target times, whether in unstratified \citep{Foures2014,HC22} or stratified fluids \citep{MC18}. Mix-norms thus provide a computationally efficient way of identifying the long-term, physical mixing processes while working with affordable target times.

In the present case, this difficulty was overcome by considering an alternative `more physics-based' cost functional such that magnetic energy of the axisymmetric component of the magnetic field is maximised. Choosing this cost functional appears not only to bias the optimisation procedure toward local extrema that do systematically trigger a sustained subcritical dynamo at long times;  it is also found more robust in the sense that repeated optimisations consistently identified  the same dynamo seed, which is also found not to significantly vary as the target time is increased.

Moreover, the identification of travelling waves solutions using a Newton-hookstep algorithm \citep{Willis_2019} confirms that the evolution initiated by these optimal dynamo seeds brings the system close to an unstable magnetohydrodynamic solution within the time horizon we used for the computation of the minimal seed. From that unstable state, the fate of the system is determined by the energy budget allowed for the initial seeds, travelling away toward either the dynamo attractor (which also corresponds to the linearly stable, magnetised travelling wave solution identified by the Newton-hookstep algorithm) or a non-dynamo one (the linearly stable, purely hydrodynamic travelling wave solution).

Therefore, we can now form a picture of the dynamical landscape of the various basins of attraction at our working parameters, which summarises the findings of the present paper. Figure \ref{fig:landscape} shows the trajectory initiated by different initial conditions in the kinetic-magnetic energy phase space, along with the travelling waves solutions (hydrodynamic TW0, stable MHD TW1, unstable MHD TW2). Whereas the \cite{Christensen2001} benchmark initial condition (rescaled here to the largest energy at which it can no longer trigger a dynamo) approaches the unstable edge state TW1 in a (somewhat) inefficient way, undergoing repeated energy transfers between the kinetic, magnetic and thermal energy reservoirs, the less energetic, minimal dynamo seed yields an important amplification of the magnetic energy by maintaining a stronger convection at early times, while energy transfers at later times are largely suppressed. This strategy rapidly propels the system to the vicinity of the edge state, from which it is repelled away towards the dynamo attractor or (when the initial energy budget is too small) the purely hydrodynamic one.

We conclude by returning to our motivation. What is the relevance of our approach for studies of the geodynamo? Finding minimal seeds and exact nonlinear solutions will be important in promoting our understanding in a field where it is impossible to perform numerical calculations in the correct parameter regime. We intend to build on this work by investigating how the minimal seeds and exact nonlinear solutions change as the underlying parameters are varied. For example, how does the minimal energy of the magnetic field perturbation change as one varies the parameters along the dynamo path \citep{agf2017} taking one to the distinguished limit of \citet{dormy_2016}. Does the energy required to take one to a fully nonlinear strong field solution decrease as the Ekman number is decreased; this behaviour would be reminiscent of the behaviour with Reynolds number of the energy of the minimal seed in wall-bounded shear flows \citep[see e.g.][]{duguet2013}. If this were the case then even a tiny magnetic energy perturbation would be enough to recover the dynamo if it entered a non-dynamo state --- even if the hydrodynamic solution were formally linearly stable. Furthermore, tracing the behaviour of exact solutions as parameters are varied is a very efficient way of examining whether (admittedly specialised) solutions could maintain balance as parameters are varied. The bifurcation structure of these solutions (particularly the presence or absence of global bifurcations) may also give hints about mechanisms for achieving reversal of solutions whilst maintaining the correct balance.

\section*{Acknowledgements}
CSS would like to acknowledge Jacob Page for helpful discussions on the Newton-hookstep algorithm. The authors thank Yohann Duguet for providing valuable feedback on the manuscript. This work was undertaken on ARC4, part of the High Performance Computing facilities at the University of Leeds, UK.

\section*{Funding}
CSS and SMT acknowledge partial support from a grant from the Simons Foundation (Grant No. 662962, GF). CSS and SMT would also like to acknowledge support of funding from the European Union Horizon 2020 research and innovation programme (grant agreement no. D5S-DLV-786780). FM acknowledges support from the  European Research Council under Grant Agreement CIRCE 101117412.

\section*{Declaration of interests}
The authors report no conflict of interest.


\appendix
\section{Derivation of the adjoint equations}\label{sec:adj}

\subsection{Adjoints of vector calculus operations}
In order to find the adjoint of our equations, we first describe how to take the adjoints of common vector calculus operations. Vector calculus gives $\vect{y}^\dagger\cdot\nabla y=\nabla \cdot (\vect{y}^\dagger y) - \vect{y}\nabla\cdot \vect{y}^\dagger$, hence
\begin{eqnarray}
\iiint \vect{y}^\dagger\cdot\nabla y \;\textrm{d}V &=& \oiint (\vect{y}^\dagger y)\cdot \vect{\textbf{\textrm{d}S}} -\iiint \vect{y}\nabla\cdot \vect{y}^\dagger \;\textrm{d}V,\\
\langle \vect{y}^\dagger,\nabla y\rangle &=& \langle -\nabla\cdot\vect{y}^\dagger,y\rangle + \{\vect{y}^\dagger y\},
\end{eqnarray}
where we have used the divergence theorem. Note that we have also introduced the notation $\{\cdot\}$ to represent the boundary terms. This identity gives the adjoint for both the gradient and divergence operators. For the curl we can use the identity $\nabla \cdot(\vect{y}^\dagger\times\vect{y})=\vect{y}\cdot \nabla\times\vect{y}^\dagger-\vect{y}^\dagger\cdot \nabla\times\vect{y}$ to obtain
\begin{equation}
   \langle\vect{y}^\dagger, \nabla\times\vect{y}\rangle=\langle\vect{y}, \nabla\times\vect{y}^\dagger\rangle - \{\vect{y}^\dagger\times\vect{y}\}.
\end{equation}
To find the adjoint of the scalar Laplacian we can use our results for the divergence and gradient to obtain
\begin{equation}
\langle y^\dagger,\nabla^2 y\rangle
=\langle \nabla^2 y^\dagger,y\rangle + \{y^\dagger\nabla y\} - \{y\nabla y^\dagger\}.
\end{equation}
The adjoint of the vector Laplacian can also be found from the rules we have derived so far, by writing $\nabla^2 \vect{y}=\nabla (\nabla\cdot\vect{y})-\nabla\times(\nabla\times\vect{y})$, yielding
\begin{equation}
\langle \vect{y}^\dagger,\nabla^2 \vect{y}\rangle =\langle \nabla^2\vect{y}^\dagger,\vect{y}\rangle + \{\vect{y}^\dagger\nabla\cdot\vect{y}\}
 -\{\vect{y}\nabla\cdot\vect{y}^\dagger\}+
 \{\vect{y}^\dagger\times(\nabla\times\vect{y})\} + \{(\nabla\times\vect{y}^\dagger)\times\vect{y}\}.
\end{equation}
These results are summarised in table \ref{tab:adj}.

\begin{table}
\begin{center}
\begin{tabular}{ c c c } 
Term & Adjoint & Boundary term\\
 \hline
 $\langle \vect{y}^\dagger,\nabla^2\vect{y}\rangle$ & $\langle\nabla^2\vect{y}^\dagger,\vect{y}\rangle$ & $\vect{y}^\dagger\nabla\cdot\vect{y}-\vect{y}\nabla\cdot\vect{y}^\dagger+\vect{y}^\dagger\times(\nabla\times\vect{y})+ (\nabla\times\vect{y}^\dagger)\times\vect{y}$\\
 $\langle y^\dagger,\nabla^2 y\rangle$ & $\langle\nabla^2 y^\dagger,y\rangle$ & $y^\dagger\nabla y -y \nabla y^\dagger$ \\
 $\langle\vect{y}^\dagger,\nabla y\rangle$ & $\langle-\nabla\cdot\vect{y}^\dagger,y\rangle$ & $\vect{y}^\dagger y$\\ 
 $\langle\vect{y}^\dagger,\nabla\times\vect{y}\rangle $ & $\langle\nabla\times\vect{y}^\dagger,\vect{y}\rangle$ &  $-\vect{y}^\dagger\times\vect{y}$\\
 $\langle\vect{y}^\dagger,\vect{a}\times\vect{y}\rangle$ & $\langle-\vect{a}\times\vect{y}^\dagger,\vect{y}\rangle$ & 0\\ 
 \hline
\end{tabular}
\end{center}
\caption{Table of terms, their corresponding adjoints, as well as boundary contributions. Effectively each row of this table can be read as the result of integration by parts. Adjoints for combinations of these operations can be obtained by applying these rules sequentially.}
\label{tab:adj}
\end{table}

\subsection{The adjoint equations}\label{sec:adjoint_derivation}
With the adjoints of general vector calculus operations now derived, we can proceed to find the adjoint our equations by cancelling all first variations of the Lagrangian 
(\ref{equ:Lagrangian}). Taking the first variation with respect to the adjoint state ensures that the governing equations are satisfied. The first variation with respect to the state $\vect{q}$, gives
\begin{equation}
    \nabla_{\vect{q}}\mathcal{L}\cdot \delta\vect{q}=\left[\frac{\partial\mathcal{J}_I}{\partial \vect{q}},\delta\vect{q}\right]+\left[\matr{M}^\mathrm{T}\frac{\partial \vect{q}^\dagger}{\partial t},\delta\vect{q}\right]+\left[\vect{q}^\dagger,\frac{\partial\boldsymbol{\mathcal{N}}(\vect{q})}{\partial \vect{q}}\delta\vect{q}\right],
    \label{equ:adjStep1}
\end{equation}
where integration by parts has been used in time. By finding the adjoint operator $\boldsymbol{\mathcal{A}}^\dagger$ such that
\begin{equation}
    \left\langle \vect{y}^\dagger,\frac{\partial\boldsymbol{\mathcal{N}}(\vect{q})}{\partial \vect{q}}\vect{y}\right\rangle= \left\langle \boldsymbol{\mathcal{A}}^\dagger\vect{y}^\dagger,\vect{y}\right\rangle,
    \label{equ:adjoint_relation}
\end{equation}
we can then rearrange (\ref{equ:adjStep1}) to 
\begin{equation}
    \nabla_{\vect{q}}\mathcal{L}\cdot \delta\vect{q}=\left[\matr{M}^\mathrm{T}\frac{\partial \vect{q}^\dagger}{\partial t}+\boldsymbol{\mathcal{A}}^\dagger\vect{q}^\dagger+\frac{\partial\mathcal{J}_I}{\partial \vect{q}},\delta\vect{q}\right].
\end{equation}
Requiring this term to be zero is achieved by setting the adjoint to satisfy
\begin{equation}
    -\matr{M}^\mathrm{T}\frac{\partial \vect{q}^\dagger}{\partial t}=\boldsymbol{\mathcal{A}}^\dagger\vect{q}^\dagger+\frac{\partial\mathcal{J}_I}{\partial \vect{q}}.
    \label{equ:adjEquationGeneral}
\end{equation}

In order to find the adjoint operator $\boldsymbol{\mathcal{A}}^\dagger$, equation (\ref{equ:adjoint_relation}) shows us that we must first linearise the governing equations. Integration-by-parts using the rules in table \ref{tab:adj} then yields the adjoint operator. For our equations, this is done using the steps outlined by \citet{chen_herreman_li_livermore_luo_jackson_2018}, where the governing equations are rewritten in a form in which Ohm's law (where Ampère's law has been used to relate the magnetic field to the current) 
\begin{equation}
\sigma\vect{E} = -\vect{U}\times\vect{B}+\frac{1}{\textit{Pm}}\nabla\times\vect{B},
\end{equation}
and Faraday's law
\begin{equation}
\frac{\partial\vect{B}}{\partial t}=-\nabla\times\vect{E},
\end{equation}
appear separately. In other words, equations for the electric field $\vect{E}$ and magnetic field are constrained separately using adjoint variables $\vect{E}^\dagger$ and $\vect{b}^\dagger$, respectively. As in \citet{chen_herreman_li_livermore_luo_jackson_2018}, $\sigma$ is a relative electrical conductivity which takes the value $1$ for $\vect{x}\in\mathcal{V}_s$ and $0$ for $\vect{x}\in\mathbb{R}^3\backslash \mathcal{V}_s$. This allows the adjoint formulation to more easily incorporate electrically insulating inner and outer regions, where the insulating boundary condition is simply continuity of the radial part of the magnetic field and tangential part of the electric field at the boundaries between $\mathcal{V}_s$ and the insulating regions. Additionally, we introduce a Lagrange multiplier $\Pi^\dagger$, which constraints the solenoidal condition $\nabla\cdot\vect{B}=0$.

Following this procedure, we obtain the adjoint equations 
\begin{equation}
\begin{gathered}
\textit{Ek}\left(-\frac{\partial \vect{u}^\dagger}{\partial t}+\nabla\times(\vect{U}\times\vect{u}^\dagger)+\vect{u}^\dagger\times\boldsymbol{\omega} - \nabla^2 u^\dagger\right) - 2 \vect{e}_z\times\vect{u}^\dagger + \nabla p^\dagger=-T^\dagger\nabla T
-\vect{B}\times\vect{E}^\dagger + \frac{\partial\mathcal{J}_I}{\partial \vect{U}},\\
\nabla\cdot \vect{u}^\dagger= 0,\\
-\frac{\partial T^\dagger}{\partial t}-\nabla\cdot(\vect{U}T^\dagger)=\tilde{\textit{Ra}}\frac{\vect{r}}{r_0}\cdot\vect{u}^\dagger+\frac{1}{\textit{Pr}}\nabla^2T^\dagger +\frac{\partial\mathcal{J}_I}{\partial T},\\
-\frac{\partial \vect{b}^\dagger}{\partial t}=-\nabla\Pi^\dagger+\frac{1}{\textit{Pm}}\nabla\times(\vect{B}\times\vect{u}^\dagger)-\frac{1}{\textit{Pm}}(\nabla\times\vect{B})\times\vect{u}^\dagger + \vect{U}\times\vect{E}^\dagger+\frac{1}{\textit{Pm}}\nabla\times\vect{E}^\dagger +\frac{\partial\mathcal{J}_I}{\partial \vect{B}},\\
\vect{E}^\dagger = -\nabla\times\vect{b}^\dagger,\\
\end{gathered}
\end{equation}
for $\vect{x}\in \mathcal{V}_s$, and 
\begin{equation}
\begin{gathered}
\nabla\times\vect{b}^\dagger = 0,
\end{gathered}
\end{equation}
for $\vect{x}\in \mathcal{V}_i\cup\mathcal{V}_o$. Alternatively, using the fact that $\vect{E}^\dagger = -\nabla\times\vect{b}^\dagger$ for $\vect{x}\in \mathcal{V}_s$ gives the form used in (\ref{equ:adjointEqus}).

These equations match the form of the equations contained in \citep{Mannix2022a}, and also \citep{chen_herreman_li_livermore_luo_jackson_2018,Holdenried-Chernoff2019} if only the induction equation with a specified velocity field is considered, but with extra terms stemming from the temperature variable. We see that the adjoint equations currently have more unknowns than equations due to the inclusion of the $\nabla\Pi^\dagger$ term in the adjoint magnetic field equation. As in the study of \citet{chen_herreman_li_livermore_luo_jackson_2018}, we see that the adjoint equations imply a gauge freedom for the adjoint magnetic field $\vect{b}^\dagger$. As it is only the curl of the adjoint magnetic field that occurs, setting $\vect{b}^\dagger\rightarrow \vect{b}^\dagger + \nabla \chi$ does not change the equations. This gauge freedom is a consequence of the direct equations being overdetermined for $\vect{B}$, with the induction equation naturally preserving the divergence of $\vect{B}$ which is physically always initialised with a divergence free field. It turns out to be convenient to then fix this gauge freedom by choosing $\vect{b}^\dagger$ to be solonoidal, i.e. we solve (\ref{equ:adjointEqus}) together with $\nabla\cdot\vect{b}^\dagger=0$, which determines the value of $\Pi^\dagger$. This gauge choice has two main benefits; firstly that the update condition (\ref{equ:updateGen}) is now guaranteed to give an updated magnetic field that is divergence free, and secondly the boundary conditions for the adjoint magnetic field will be the same as for the direct magnetic field (see \citet{chen_herreman_li_livermore_luo_jackson_2018}, for example). Cancelling the boundary terms that arise from integration-by-parts of terms involving the velocity and temperature fields gives the adjoint boundary conditions.

\section{Cost functionals}\label{sec:cost_functionals}
For each cost functional considered we need to be able to find its contribution to the adjoint equation. For optimising the total magnetic energy (\ref{equ:JI_int}), we obtain
\begin{eqnarray}
    \frac{\partial \mathcal{J}_I}{\partial \vect{B}} = \frac{\vect{B}}{\textit{Ek}\textit{Pm}}.
\end{eqnarray}
Similarly, for optimising the axisymmetric energy (\ref{equ:JI_m_int})
we find
\begin{eqnarray}
    \frac{\partial \mathcal{J}_I}{\partial \vect{B}} = \frac{\bar{\vect{B}}}{\textit{Ek}\textit{Pm}}.
\end{eqnarray}

\section{Convergence}\label{sec:convergence}

\begin{figure}
  \centering
  \begin{tabular}{ccc}
  \begin{subfigure}[t]{0.45\textwidth}
      \includegraphics[width=\linewidth]{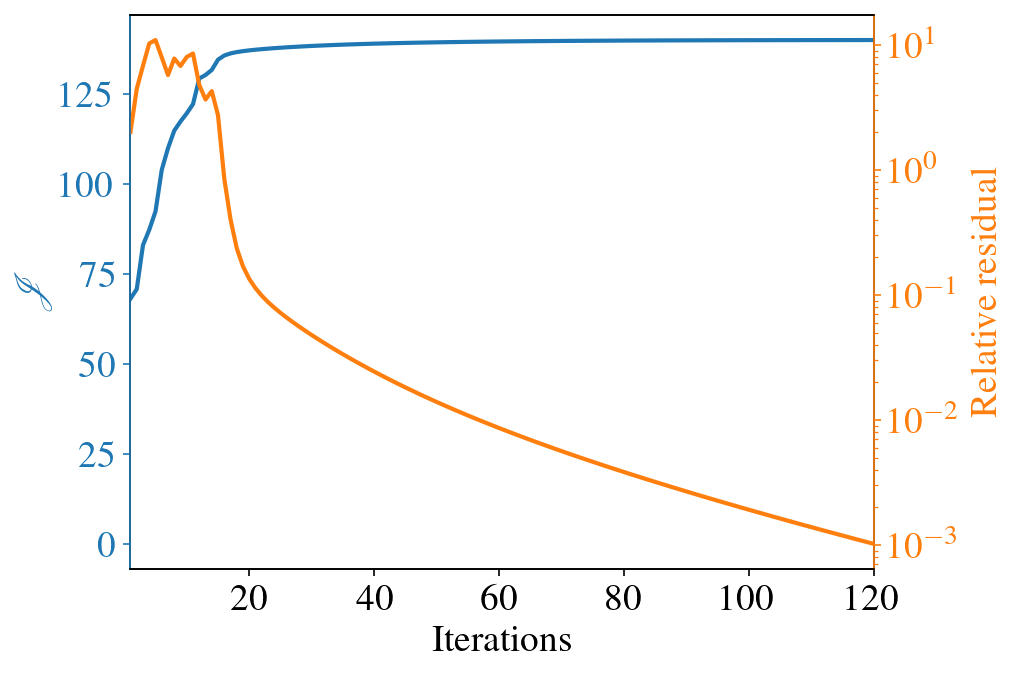}
      \caption{Total energy cost functional.}
    \end{subfigure}
    &
    \begin{subfigure}[t]{0.45\textwidth}
      \includegraphics[width=\linewidth]{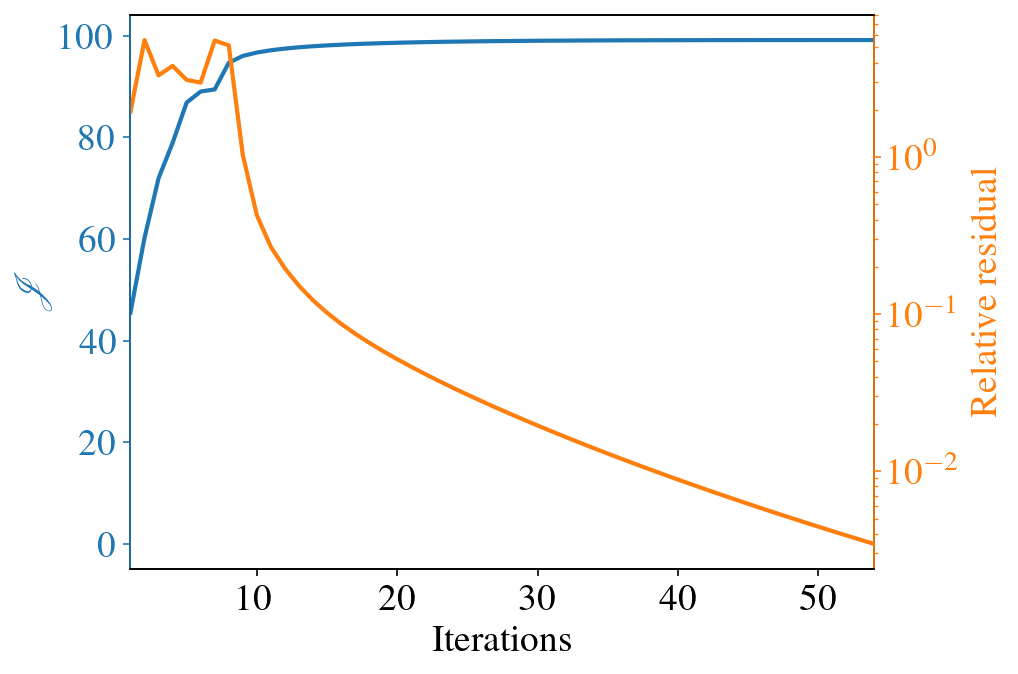}
      \caption{Axisymmetric energy cost functional.}
    \end{subfigure} 
  \end{tabular}
  \caption{Convergence behaviour for the optimisation procedure described in \S\ref{sec:opt} with two different cost functional with $t_\textrm{opt}=0.2$ and $\textit{M}_0=344$ --- and with $l=m=2$ initially.}
  \label{fig:convergence}
\end{figure}

The typical convergence behaviour of the optimisation procedure is illustrated in figure \ref{fig:convergence}, which shows the evolution of both the cost functional and the relative residual over the iterations. We have defined the residual as
\begin{equation}
r = \frac{\iiint_{\mathcal{V}_s} \vect{g}_\perp\cdot\vect{g}_\perp \;\textrm{d}V}{\mathcal{J}},
\end{equation}
where $\vect{g}_\perp$ is the Riemannian gradient that does not change the norm of $\vect{B}_0$. The cost functional rapidly increases at first, as the large scale structure of the initial magnetic field undergoes rapid changes from its random initial guess. We perform the optimisation procedure until either the residual is less than $10^{-3}$ or 120 iterations have occurred. After approximately 20 iterations the cost functional tends toward a constant value as minor, mostly small-scale adjustments occur, while the relative residual steadily decreases.

\end{document}